\documentclass[fleqn,usenatbib]{mnras}
\usepackage[british]{babel}
\usepackage{newtxtext,newtxmath}
\usepackage[T1]{fontenc}

\DeclareRobustCommand{\VAN}[3]{#2}
\let\VANthebibliography\thebibliography
\def\thebibliography{\DeclareRobustCommand{\VAN}[3]{##3}\VANthebibliography}

\usepackage[normalem]{ulem}
\usepackage{graphicx}	% Including figure files
\usepackage{amsmath}	% Advanced maths commands
\usepackage{amssymb}	% Extra maths symbols
\usepackage{multirow}
\usepackage{mathtools}

\usepackage{dcolumn}
\usepackage{siunitx}
\newcommand{\linka}[1]{\href{#1}{\url{#1}}}
\let\urshowkeys=\showkeys \def\showkeys{\needspace{5ex}\urshowkeys}
\newcommand{\Msun}{\ensuremath{M_{\odot}}}

\newcommand{\Log}{\mathrm{ln}}
\renewcommand{\ln}{\Log}
\newcommand{\vir}{\texttt{vir}}

\newcommand{\grassetto}[1]{#1}
\newcommand{\juno}[1]{{ #1}}
\newcommand{\junof}[1]{\juno{  #1}}
\newcommand{\dd}{{\Delta2}}
\newcommand{\du}{{\Delta1}}

\title[$c_{\Delta}\left(M_{\Delta}, z, \rm{cosmology}\right)$]{Cosmology dependency of halo masses and concentrations  in  hydrodynamic simulations}
\author[A. Ragagnin]{Antonio Ragagnin$^{1,2}$, Alexandro Saro$^{3,2,1,4}$, Priyanka Singh$^{1,2},$ Klaus Dolag$^{5,6}$
\\
$^{1}$ INAF-Osservatorio Astronomico di Trieste, via G.B. Tiepolo 11, 34143 Trieste, Italy\\
$^{2}$ IFPU - Institute for Fundamental Physics of the Universe, Via Beirut 2, 34014 Trieste, Italy\\
$^{3}$ Astronomy Unit, Department of Physics, University of Trieste, via Tiepolo 11, I-34131 Trieste, Italy \\
$^4$ INFN - National Institute for Nuclear Physics, Via Valerio 2, I-34127 Trieste, Italy \\
$^5$ University Observatory Munich, Scheinerstr 1, D-81679 Munich, Germany\\
$^6$ Max-Planck-Institut f\"{u}r  Astrophysik (MPA), Karl-Schwarzschild Strasse 1, 85748 Garching bei M\"{u}nchen, Germany\\}
 
\pubyear{2020}

\begin{document}
\label{firstpage}
\pagerange{\pageref{firstpage}--\pageref{lastpage}}
\maketitle

\begin{abstract}
We employ a set of  Magneticum cosmological hydrodynamic simulations that span over $15$ different cosmologies, and extract masses and concentrations of all well-resolved haloes between $z=0-1$ for critical over-densities  $\Delta_\texttt{vir}, \Delta_{200c}, \Delta_{500c}, \Delta_{2500c}$ and mean overdensity $\Delta_{200m}.$
\juno{We % observational surveys
% a correct mass-calibration of a total matter profiles needs a mass-concentration relation from hydrodynamic simulations and  
%first show how a full physics simulation produces haloes $10\%-20\%$ less concentrated than non-radiative runs, which motivates us to
provide the first mass-concentration (Mc) relation and sparsity relation (i.e. $M_\du -M_\dd$ mass conversion) of  hydrodynamic simulations that is modelled by mass, redshift and cosmological parameters $\Omega_m, \Omega_b, \sigma_8, h_0$   as a tool for observational studies. }
\juno{We also quantify the impact that the Mc relation scatter  and the assumption of NFW density profiles have on the uncertainty of the sparsity relation.}
We find that converting masses with the aid of a Mc relation carries an additional  fractional scatter ($\approx 4\%$) originated from deviations from the assumed NFW density profile. 
\grassetto{
For this reason we provide a direct mass-mass conversion relation fit  that depends on redshift and cosmological parameters.
}
We release the package \texttt{hydro\_mc}, %\protect\footnote{\href{https://github.com/aragagnin/hydro_mc}{github.com/aragagnin/hydro\_mc}}
a python tool that perform all kind of conversions presented in this paper.
\end{abstract}

\begin{keywords}halo - large-scale structure of Universe - cosmological parameters\end{keywords}

\section{Introduction}

Early studies of numerical N-body simulations of cosmic structures embedded in cosmological volumes \citep[see e.g.][]{1997ApJ...490..493N,1997ApJS..111...73K} showed that dark matter haloes can be described by the so called Navarro-Frank-White (NFW) profile~\citep{1996ApJ...462..563N}.
The NFW density profile $\rho\left(r\right)$ is modelled by a characteristic density  $\rho_0$ and a scale radius $r_s$ in the following way:
\begin{equation}
    \rho\left(r\right)=\cfrac{\rho_0}{\cfrac{r}{r_s}\left(1+\cfrac{r}{r_s}\right)^2}.
    \label{eq:nfw}
\end{equation}

The NFW profile proved to match  density profiles of  dark matter haloes of  dark-matter-only (DMO) simulations \citep[see e.g.][]{2001MNRAS.321..559B,2003astro.ph.11575S,2012MNRAS.423.3018P,2014ApJ...797...34M,2016MNRAS.457.4340K,2017MNRAS.469.3069G,2019AAS...23333806B} up to  the  largest and most resolved ones whose analyses trace the route  for the next generation of (pre-)Exascale simulations.
However, density profiles of hydrodynamic simulations have small deviations from the NFW profile  \citep[see e.g.][]{2014MNRAS.437.2328B,2016MNRAS.456.3542T}.

Since this kind of density profile does not have a cut-off radius, the radius of a halo is often chosen as the   virial radius $R_{\texttt vir}$~\citep[see e.g.][]{1998MNRAS.300..146G,1999ApJ...525..554F},
namely, the radius at which the mean density  crosses the one of a theoretical virialised homogeneous top-hat overdensity.
\cite{1998ApJ...495...80B} showed that the virial overdensity can be written as
\begin{equation}
\Delta_{\vir}(a) \approx 18\pi^2  + 82\Omega\left(a\right) - 39\Omega\left(a\right),
    \label{eq:delta}
\end{equation}
where $a$ is the scale factor and $\Omega(a)$ is the energy density parameter \citep[see][for a review]{2003moco.book.....D}, namely
\begin{equation}\Omega(a) = \Omega_m  a^3 \times \left(\frac{\Omega_m}{a^3}  + \frac{\Omega_r}{a^4} + \frac{ \Omega_k} {a^2} + \Omega_\Lambda\right)^{-1},\label{eq:omegas}
\end{equation}
where $\Omega_m, \Omega_r, \Omega_k$ and $\Omega_\Lambda$
are the density fractions of the total matter, radiation, curvature and cosmological constant, respectively. Numerical cosmological simulations, as in this work, typically use negligible radiation and curvature terms (they set $\Omega_r=\Omega_k=0$ in Eq. \ref{eq:omegas}).

%Modern theoretical studies in the form of cosmological simulations, can be subdivided into dark matter only simulations \citep{2012MNRAS.426.2046A,2012MNRAS.423.3018P}, hydrodynamic non-radiative simulations (where lack of radiation forbids gas to cool and form stars\citep{2016MNRAS.457.4063S}), and simulations where all major phenomena are included, as cooling, star formation, black hole seeding and their consequent feedback \citep[see e.g.][]{2010MNRAS.405.2161D}.

Observational studies typically define galaxy cluster (GC) radii as $R_{\Delta c},$ where $\Delta$ is an arbitrary overdensity and the "$c$" suffix indicates that the overdensity is relative to the critical overdensity given by,
\begin{equation}
%M(r<R_{\Delta c}) = \int_{\vec r<R_{\Delta c}} d\vec r \rho(r) =  \cfrac{4}{3}\pi R_{\Delta c}^3\cdot \Delta\cdot\rho_c.
M(r<R_{\Delta c}) =  \Delta \times \cfrac{4}{3}\pi R_{\Delta c}^3 \rho_c.
\label{eq:mdelta}
\end{equation}
X-ray observations typically use overdensities $\Delta_{500c}$ and $\Delta_{2500c}$ and the corresponding radii $R_{500c}$ and $R_{2500c}$ \citep[see e.g.][]{2019ApJ...878...55B,2019arXiv190910524U,2019MNRAS.485.4863M,2019ApJ...871...50B}, whereas,
%(the mean density crosses   $500\rho_c$ and $2500\rho_c$ respectively), see e.g. \cite{2019ApJ...878...55B,2019arXiv190910524U,2019MNRAS.485.4863M,2019ApJ...871...50B}.
observational studies that compute dynamical masses typically use  $\Delta=\Delta_{200c}$  \citep[see e.g.][]{2017A&A...607A..81B,2019arXiv191004773C}.
Weak Lensing  studies on the other hand often utilise radii whose overdensities are proportional to the mean density of the Universe.
For instance, works such as \cite{2008JCAP...08..006M,2019MNRAS.482.1352McClintock}  measure halo radii as $R_{200m},$ where the suffix "$m$"  means that the radius is defined so the mean density of the halo  in Eq. \ref{eq:mdelta} crosses $\Delta\overline{\rho},$ (in this case $200\overline{\rho}$) where $\overline{\rho}$ is the average matter density of the Universe.

The concentration $c_{\Delta}$ of a halo is defined as, $c_{\Delta}\equiv R_{\Delta}/r_s$, where, 
%\begin{equation}
%c_{\Delta}\equiv R_{\Delta}/r_s,
%\label{eq:c} \end{equation}
$r_s$ is the scale radius of Eq. \ref{eq:nfw} and  quantifies how large  the internal region of the cluster is compared to its radius for a given overdensity~\citep[see][for a review]{2017arXiv171105277O}.
Both numerical and observational studies analyse the concentration of haloes in the context of the so called mass-concentration (Mc) plane \citep[see  Table 4 in][for comprehensive  list of recent studies]{2019MNRAS.486.4001R}. 
\juno{Within the context of hydrodynamic simulations, one can define the DM mass-concentration plane which can be used by observations that  estimate  DM profiles~\citep[e.g.][]{2015ApJ...806....4Melchior}. On the other hand observations that  have only information on the total-matter profile must rely on total-matter mass concentration planes~\citep[e.g][]{2019ApJ...872..170Raghunathan}.}

\juno{In search of a realistic estimate of halo concentrations, one must consider the various sources that affect this value.} 
The $c$ parameter in both observational and numerical studies is found to have a weak dependence on halo mass and a very large scatter~\citep{2001MNRAS.321..559B,2013A&A...557A.131M,2014MNRAS.441..378L,2017ApJ...840..104S,2018MNRAS.477.2804S,2019MNRAS.486.4001R}. Concentration has been found to depend on a number of factors, as formation time of haloes~\citep{2001MNRAS.321..559B,2018arXiv181009473R},   accretion histories~\citep[see e.g.][]{2013MNRAS.432.1103L,2018ApJ...857..118F,2018ApJ...863...37F}, dynamical state \citep{2012MNRAS.427.1322L}, triaxiality \citep{2012MNRAS.426.1558G,2014MNRAS.440.1899G}, and halo environment~\citep{2018A&A...618A.172C,2016MNRAS.457.4340K}.
The fractional scatter in the Mc plane is larger than $\gtrsim 33\%$ \citep{2016ApJ...820..108H}, and observations found outliers with  extremely high concentration \citep{2019ApJ...877...91B} or very low concentration \citep{2019A&A...630A..78A}.
 When all major physical phenomena of galaxy formation  are taken into account (cooling, star formation, black hole seeding and their feedback), then concentration parameters are lower than their dark-matter-only counterpart \citep[see e.g. results from NIHAO simulations as in][]{2015MNRAS.454...83W,2016MNRAS.456.3542T}.

%{\ccr Is there something missing in the first sentence?}
\juno{Halo concentration parameters are also affected by the underlying  cosmological model~\citep[see e.g.][for a review on cosmological models]{2003icte.book.....R}. The derived Mc relation is found in fact different in Cold Dark Matter (CDM), $\Lambda$CDM, $w$DM, and varying dark energy equation of state \citep[][]{1997ApJS..111...73K,2016MNRAS.460.1214L,2004A&A...416..853D,2013arXiv1302.2364D,2013MNRAS.428.2921D}.
In general, the Mc dependency of DMO simulations on cosmological parameters has been extensively studied in works as the Cosmic Emulator \citep{2008MNRAS.391.1940M,2013ApJ...766...32B,2016ApJ...820..108H}, works as \cite{2014MNRAS.441..378L} and \cite{2012MNRAS.423.3018P}.}

%\juno{From observational perspective,} mass-concentration plane is an important tool to test cosmological models \citep{2019arXiv190802508K} and to convert masses between two over-densities.
\juno{Mass-concentration relations allow observational works to convert masses between overdensities.}
For this purpose, \cite{2014MNRAS.437.2328B} defined the sparsity parameter  $s_{\du,\dd}$ as the ratio between masses at  over-density $\Delta_1$ and $\Delta_2$.
This quantity is a proxy to the total matter profile \citep{2018ApJ...862...40C} and enables cosmological parameter inference \citep{2019MNRAS.487.4382C} and testing for some dark energy models without assuming an NFW profile \citep{2014MNRAS.437.2328B}.
Observations use the sparsity parameter to infer the halo matter profile \citep{2019A&A...628A..86B}, as a potential probe to test $f(R)$ models \citep{2016PhRvD..93j3522A}, a less uncertain measurement of the mass-concentration relation \citep{2019Galax...7....8F}, and to find outliers in scaling relations involving integrated quantities with different radial dependencies \citep[see conclusions in][]{2019A&A...630A..78A}.

{%\ccr 
\juno{In this work we use data from the Magneticum suite of simulations \citep[presented in works such as][]{2013MNRAS.428.1395B,
2014MNRAS.440.2610S, 2015MNRAS.448.1504S,  2015ApJ...812...29T, 2015MNRAS.451.4277D, 
2016MNRAS.463.1797D,
2016MNRAS.458.1013S, 2016MNRAS.456.2361B, 2017MNRAS.464.3742R} to calibrate the cosmology dependence of the mass-concentration and of the mass-sparsity relation of the total matter component from hydrodynamic simulations with the purpose of facilitating cluster-cosmology oriented studies.}
\juno{These studies typically  calibrate the observable-mass relation from stacked weak lensing signal under the assumption that mass-calibration can be correctly recovered from DMO Mc relations (e.g.
\citealt{2014MNRAS.438...78Rozo};
\citealt{2014MNRAS.443.1713Dietrich};
\citealt{2016MNRAS.461.4099Baxter};
\citealt{2017MNRAS.466.3103Simet};
\citealt{2017NatAs...1..795Geach};
\citealt{2019MNRAS.482.1352McClintock};
\citealt{2019ApJ...872..170Raghunathan} and references therein), an approximation that has to be quantified by calibrating the total-mass  mass-concentration relations within hydrodynamic simulations~\citep[see discussion in Sec. 5.4.1 of][]{2019MNRAS.482.1352McClintock}}. 
}

\junof{This work represents a first necessary step in this direction and it provides  mass-concentration and mass-mass relations that depends on cosmology and that simultaneously accounts for the presence of baryons. 
While in fact previous works in the literature studied either the dependency of the concentration on cosmological parameters or on baryon physics, in this analysis we calibrate for the first time the dependency of concentration on cosmological parameters in the context of hydrodynamic simulations that include a full description of the main baryonic physical processes.}

In Section \ref{sec:sims} we present the numerical set up of the simulations used in this work.
In Section \ref{sec:mca} we fit the concentration of haloes as a function of mass and scale factor for all our simulations and compare our results with both observations and other theoretical studies.
In Section \ref{sec:mcac}  we provide a fit of the concentration as a function of mass, scale factor and cosmology.
\juno{As uncertainty propagation is a delicate and important matter for cluster cosmology experiments, in Sec. \ref{sec:mmac} we test sparsity parameter and study the origin of its large uncertainty}.
%{\ccr{In order to facilitate cluster cosmology studies that include mass-observable relations which are calibrated at different radii (e.g. bocuet15,19,mantz18, costanzi20, etc..), we study how to convert masses at different overdensities (the sparcity-mass relation). We summarize our findings, including a careful characterization of the associated intrinsic scatter. in Sec. \ref{sec:mmac}}}.
\juno{In order to facilitate cluster cosmology studies that include mass-observable relations which are calibrated at different radii \citep[e.g.][]{2016MNRAS.456.2361B,2019ApJ...878...55B,2019MNRAS.485.4863M,2019MNRAS.488.4779Costanzi}, we study how to convert masses at different overdensities (the sparsity-mass relation). We summarise our findings, including a careful characterisation of the associated intrinsic scatter. in Sec. \ref{sec:mmac}}.
We draw our conclusions in Section \ref{sec:conclu}.

%%%%%%%%%%%%%%%%%%%%%%%%%%%%%%%%%%%%%%%%%%%%%%%%%%%%%%
%%%%%%%%%%%%% Numerical Simulations %%%%%%%%%%%%%%%%%%
%%%%%%%%%%%%%%%%%%%%%%%%%%%%%%%%%%%%%%%%%%%%%%%%%%%%%%

\section{Numerical Simulations}

\label{sec:sims}

%
% tbl:sims
%
\begin{table*}
    \caption{List of Magneticum simulations as presented in    \protect\cite{2019Singh}.     Columns show, respectively: simulation name, cosmological parameters   $\Omega_m, \Omega_b, \sigma_8,$ and   $h_0$, the number   of haloes selected from all redshift snapshots ($z=0.00,\ 0.14,\ 0.29,\ 0.47,\ 0.67,\ $ and $z=0.9$) of a given simulation and the number     of haloes of that simulations   at redshift $z=0.$ Two of these simulations  were also run  without radiative processes (C1{\_}norad and  C1{\_}norad) and C8 uses the reference cosmology from \protect\cite{2011ApJS..192...18K}.}
    \begin{tabular}{lrrrrrrl}
\hline
 Name       &   $\Omega_m$  &   $\Omega_b$ &   $\sigma_8$ &   $h_0$ &   $N_{\texttt haloes}$ &  $N_{\texttt haloes}$                       \\
        &   &    &    &   &   (all snapshots) &  (snapshot $z=0$) &                       \\
\hline
 C1         &         0.153 &       0.0408 &        0.614 &   0.666 &         29206 &        9153                        \\
 C1\_norad  &         0.153 &       0.0408 &        0.614 &   0.666 &         27613 &        9208  \\
 C2         &         0.189 &       0.0455 &        0.697 &   0.703 &         54094 &       16236                        \\
 C3         &         0.200   &       0.0415 &        0.850  &   0.730  &        107423 &       27225                        \\
 C4         &         0.204 &       0.0437 &        0.739 &   0.689 &         66351 &       19051                       \\
 C5         &         0.222 &       0.0421 &        0.793 &   0.676 &         84087 &       22037                        \\
 C6         &         0.232 &       0.0413 &        0.687 &   0.670  &         47045 &       14930                        \\
 C7         &         0.268 &       0.0449 &        0.721 &   0.699 &         58815 &       17990                        \\
 C8         &         0.272 &       0.0456 &        0.809 &   0.704 &         79417 &       22353                       \\
 C9         &         0.301 &       0.0460  &        0.824 &   0.707 &         96151 &       26473                       \\
 C10        &         0.304 &       0.0504 &        0.886 &   0.740  &        120617 &       32551                        \\
 C11        &         0.342 &       0.0462 &        0.834 &   0.708 &         97392 &       27100                        \\
 C12        &         0.363 &       0.0490  &        0.884 &   0.729 &        118342 &       33571                        \\
 C13        &         0.400   &       0.0485 &        0.650  &   0.675 &         35503 &       14626                        \\
 C14        &         0.406 &       0.0466 &        0.867 &   0.712 &        104266 &       30918                        \\
 C15        &         0.428 &       0.0492 &        0.830  &   0.732 &         92352 &       28348                        \\
 C15\_norad &         0.428 &       0.0492 &        0.830  &   0.732 &         79399 &       25270  \\
\hline
\end{tabular}
    \label{tbl:sims}
\end{table*}

Magneticum simulations are performed with an extended version of the N$-$body/SPH code P-Gadget3,
which is the successor of the code P-Gadget2 \citep{2005Natur.435..629S,2005MNRAS.364.1105S,2009MNRAS.398.1150B}, with a space-filling curve aware neighbour search \citep{2016pcre.conf..411R}, an improved Smoothed Particle Hydrodynamics (SPH)  solver \citep{2016MNRAS.455.2110B};
treatment of radiative cooling, heating, ultraviolet (UV) background,  star formation and stellar feedback processes as in \cite{2005MNRAS.361..776S} connected to a detailed chemical evolution and enrichment model as in \cite{10.1111/j.1365-2966.2007.12070.x}, which follows  11 chemical  elements (H, He,
C, N, O, Ne, Mg, Si, S, Ca, Fe) with the aid of CLOUDY photo-ionisation
code \citep{1998PASP..110..761F}.  \cite{2010MNRAS.401.1670F,2014MNRAS.442.2304H} {describe prescriptions for black hole growth and for feedback from AGNs}.

Haloes are identified using  the version of SUBFIND \citep{2001MNRAS.328..726S}, adapted by  \cite{2009MNRAS.399..497D} to take the baryon component into account.

\juno{Magneticum sub-grid physics does reproduce realistic haloes\footnote{
In particular, Magneticum simulations match observations of angular momentum  for different morphologies \citep{2015ApJ...812...29T,2016ilgp.confE..41T}; the  mass-size relation
 \citep{2016ilgp.confE..43R,2017MNRAS.464.3742R,2019MNRAS.484..869V};
the dark matter fraction \citep[see Figure 3 in][]{2017MNRAS.464.3742R};
the  baryon conversion efficiency  \citep[see Figure 10 in][]{2015MNRAS.448.1504S};   kinematical observations of early-type galaxies  \citep{2018MNRAS.480.4636S};  the  inner slope of the total matter density profile  \cite[see Figure 7 in][]{2018MNRAS.476.4543B}, the ellipticity and velocity over
velocity dispersion ratio  \citep{2019MNRAS.484..869V}; and reproduce the high concentration of high luminosity gap of  fossil objects \citep{2019MNRAS.486.4001R}.}, thus one can assume that its concentration parameter are realistic and of general applicability for purposes of calibration on observational studies. }

Table \ref{tbl:sims} gives an overview of the cosmological simulations used in this work.
They have  already been presented in \cite{2019Singh} (see Table 1 in their paper) and labelled as C1--15. Each simulation covers a volume of $896 Mpc/h,$  gas and DM particle masses respectively equal to $m_{gas}=2.6\cdot10^{9}\Msun/h$ and $m_{DM}=1.3\cdot10^{10}\Msun/h$ and a softening $\epsilon=10$ in units of comoving kiloparsec. 
They have different cosmological parameters $\Omega_m, \Omega_b, h,$ and $\sigma_8,$ exception for two simulations with the same setup as C1 and C15 (C1\_norad and C15\_norad)  that have been run without radiative cooling and star formation.

For each simulation we study the haloes at a timeslice with redshifts $z=0.00,\ 0.14,\ 0.29,\ 0.47,\ 0.67,\ $ and $z=0.90.$
In the following sections we repeat the same analyses for  overdensities $\Delta_{\texttt{vir}}, \Delta_{200c}, \Delta_{500c}, \Delta_{2500c}, \Delta_{200m}$ performing a corresponding mass-cut  (respectively on  $M=M_\texttt{vir}, M_{200c}, M_{500c}, M_{2500c}, M_{200m}$) that ensures that all haloes have at least $10^{4}$ particles. This cut is different for each of our simulations. This is opposed to what was used in \cite{2019Singh}, where they choose a fixed mass cut for all C1-C15 simulations. 
\grassetto{
The mass range of these haloes is between $10^{14} -4\times10^{15} M_{\odot}$, which fits the typical  range of galaxy cluster weak-lensing masses  \citep[see e.g.][]{2014MNRAS.439...48A}.
}

In this work we fit the NFW profile (see Eq. \ref{eq:nfw}) over the \textit{total matter component} (i.e. dark matter and baryons) as opposed to previous works \citep[see][]{2019MNRAS.486.4001R} where the NFW profile  fit was performed over the dark matter component only.
We   fit  the density profile over $20$ logarithmic bins, starting from $r=100kpc$ \citep[similar to the cut applied in  observational studies as][]{2019MNRAS.483.2871D}.
All fits with a $\chi^2>10^3$ have been excluded from our analyses (which accounts for a few hundred haloes per snapshot) as they correspond to heavily perturbed objects. 

Although works on simulations typically present quantities in comoving units of $h$ (e.g. distances in $a\cdot kpc/h$), unless specified, all quantities expressed in this work are in physical units and are not in units of $h.$

%%%%%%%%%%%%%%%%%%%%%%%%%%%%%%%%%%%%%%%%%%%%%%%%%%%%%%
%% %%% %% % % Halo concentrations % % %% %% %% %% %% %
%%%%%%%%%%%%%%%%%%%%%%%%%%%%%%%%%%%%%%%%%%%%%%%%%%%%%%

\section{Halo concentrations}
\label{sec:mca}
 
\junof{In Appendix \ref{ap:effects} we study the effect of baryons on mass-concentration planes  and show how an incorrect treatment of baryons can lead to under-estimation of the concentration up to $20\%$ and how the interplay between dark matter and baryons put the dynamical state of hydrodynamic simulations in a much more complex picture than the one of DMO simulations.}

\junof{This motivates us to study halo masses and concentrations on hydrodynamic simulations, and in particular we focus on their dependency on cosmological parameters}.
We  perform a fit of the concentration as a function of mass and redshift for each simulation at each over-density \junof{of Magneticum simulations}.
The functional form of the concentration is chosen as a power law on mass and scale factor as done in the observational studies \citep[see e.g.][]{2015ApJ...806....4Melchior} as:

\begin{equation}
\Log\ c_\Delta \left(M_\Delta\right) = \Log A + B \Log \left(\frac{M_\Delta}{M_p}\right) + C  \ln\left(\frac{a}{a_p}\right).
\label{eq:mcz}
\end{equation}
Here $A,B$ are fit parameters,
$a_p,M_p$ are median of mass and scale factor, respectively and are used as  pivot values,  and the fit will be performed including a logarithmic scatter $\sigma$.

We maximised the following likelihood $\mathcal{\hat L}$\footnote{we used the \texttt{python} package \texttt{emcee}~\citep{2013PASP..125..306F}} with a uniform prior  for all fit parameters:

\begin{equation}\ln\ \mathcal{\hat L} = -\frac{1}{2}\left( \ln(2\pi\sigma^2)  + \left(\frac{\ln\ c_\Delta\left(M_\Delta, A, B, C\right)-\ln\ c_\Delta}{\sigma}\right)^2\right).
\label{eq:l}
\end{equation}

\begin{figure*}
  \caption{Each panel shows the mass-concentration plane one full physics Magneticum simulation presented in Table \ref{tbl:sims}. Concentrations are computed  at overdensity $\Delta_{\texttt vir}.$  Data points represents   all selected haloes at redshift $z=0,$ colour-coded by their $log_{10}\chi^2.$ Concentration values are plotted  only in the range $c_{\vir}=1-10,$ because this range contains vast majority of haloes. Black line corresponds to the mass-concentration relation obtained by the fit in Eq. \ref{eq:mcz}. Gray lines corresponds to the mass-concentration relation obtained for the simulation C8 (which uses the reference cosmology \protect\cite{2009ApJS..180..330K}). The different mass-cut on each panel is due to   our  choice of selecting the smallest mass-cut where all haloes with at least $10^4$ particles. As a consequence, our mass-cuts depend on cosmological parameters.}
  \includegraphics[width=1.1\linewidth]{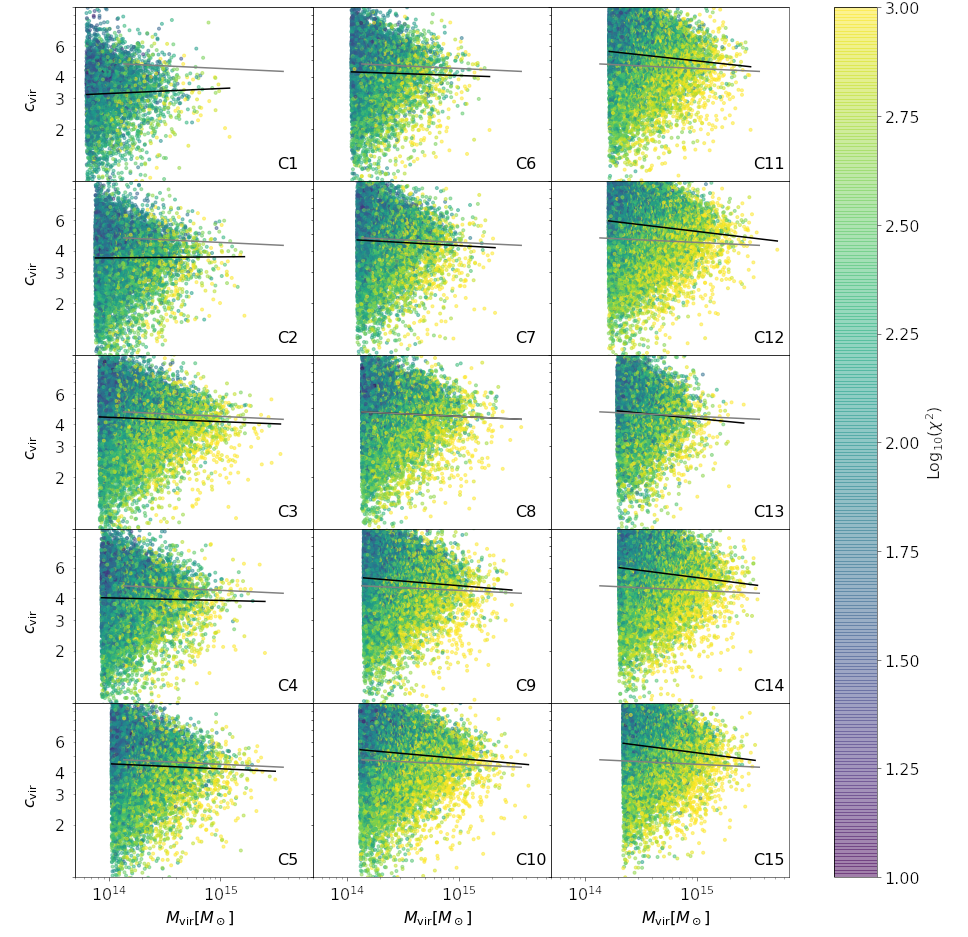}
  \label{fig:mczs}
\end{figure*}

Figure \ref{fig:mczs} shows the mass concentration planes for $\Delta_\texttt{vir}$ (computed following Eq. \ref{eq:delta}) for all $15$ simulations, together with the concentration from the redshift-mass-concentration (aMc relation) colour coded by $log_{10}\chi^2$.
Haloes with high $\chi^2$ tend to have lower concentration
which qualitatively agrees with other theoretical studies that show how perturbed objects have lower concentrations \citep[see e.g.][]{2014MNRAS.437.2328B,2014MNRAS.441..378L,2016MNRAS.457.4340K}.
For this reason, in a mass-concentration plane, it is not advisable to weight halo concentrations with  $1/\chi^2$, as this would  bias the relation towards higher concentrations.
Although the concentration is believed to decrease with increasing halo mass, extreme cosmologies such as C1 and C2 (with $\Omega_m<0.2$) have an overall positive dependency between the mass and concentration.
On the other hand, the logarithmic mean slope is low (between $-0.03$ and $0.08$) and its  influence in the mass concentration plane is not dominant in our mass regime of interest.

\subsection{Comparison with other studies}

We then compare   Magneticum simulations concentrations of haloes with the concentration predicted by  the Cosmic Emulator \citep{2016ApJ...820..108H,2013ApJ...766...32B}.
The Cosmic Emulator is a tool to predict the mass-concentration planes for a given $w$CDM. 
We were able to compare only C7, C8 and C9 cosmologies because the other Magneticum simulations had cosmological parameters that were out of the range of the Cosmic Emulator.
\junof{Note that while the Cosmic Emulator dependency on $\Omega_b$ is encoded in the power spectrum normalisation, our mass-concentration relation dependency on $\Omega_b$ takes into account all physical processes of baryon physics, including star formation and feedback.}

The ratio of median concentration $c_{\vir}$ parameters of haloes obtained with our mass-concentration fit and the concentration provided by the Cosmic Emulator is $\approx 1.2.$
We notice how the Cosmic Emulator  concentrations (retrieved by dark matter only runs) are systematically higher than Magneticum simulations in this mass regime (by a factor of $\approx 10-20\%$), in agreement with our comparison in \cite{2019MNRAS.486.4001R}.
The scatter is constant over mass, redshift and  cosmology, to nearly $\sigma\approx0.38,$ in agreement with the value of $\approx 1/3$  presented in the $w$CDM dark-matter only model of \cite{2013ApJ...768..123K}.

\begin{figure*}
  \caption{Mass-concentration plane of our simulations C1--C15 (black solid lines), haloes from the hydrodynamic cosmological simulation Omega500 \protect\citep{2018MNRAS.477.2804S} (dashed orange line), observations of fossil groups from \protect\cite{2016A&A...590L...1P} (green data points), mock observations from \protect\cite{2014ApJ...797...34M} (red data points)  and data from CLASH \protect\citep{2014ApJ...797...34M} (magenta data points) and a low-concentration halo studied in \protect\cite{2019A&A...630A..78A}. Shaded area is the scatter around the $C8$ relation.}
  \includegraphics[width=1.05\linewidth]{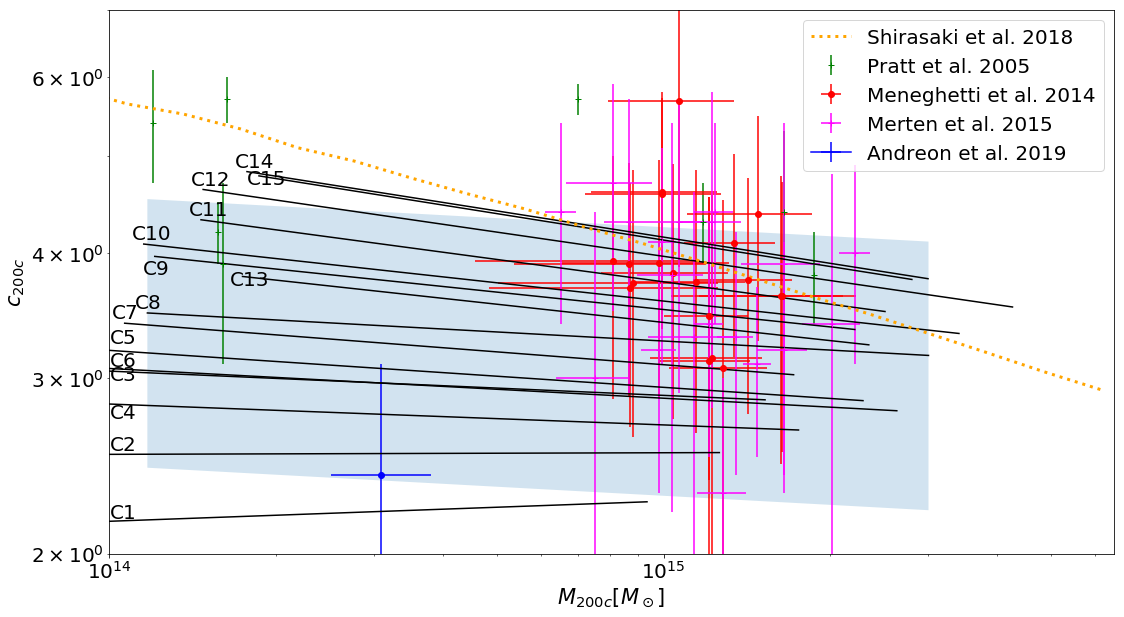}
  \label{fig:obs}
\end{figure*}

Figure \ref{fig:obs} shows the mass-concentration plane for the full-physics simulations C1--15 against other dark matter only simulations and observations.
We compare with the concentration from Omega500 simulations  \citep{2018MNRAS.477.2804S}; CLASH concentrations from \cite{2015ApJ...806....4Melchior}, numerical predictions from MUSIC of CLASH \citep{2014ApJ...797...34M}, where a number of simulated haloes have  been chosen to make mock CLASH observations.
To highlight the high scatter in the mass-concentration relation, we also show high concentration groups from \cite{2016A&A...590L...1P} and an under-luminous and low-concentration halo studied in \cite{2019A&A...630A..78A}.
 When analysing this data one must be aware of their selection effects: CLASH  data-set underwent some filtering difficult to model, while fossil  objects presented in \cite{2016A&A...590L...1P} by construction lay in the upper part of the Mc plane.
 There is a general agreement between concentration of Magneticum  simulations and these observations.

\section{Cosmology dependence of concentration parameter}
\label{sec:mcac}

The 15 cosmologies we use in this work have different mass-concentration normalisation values and log-slope  (see Figure  \ref{fig:mczs}).
we perform a fit of the concentration as  a function of mass, scale factor and cosmological parameters in order to interpolate a mass-concentration plane at a given, arbitrary, cosmology, i.e. 
a concentration $c_\Delta \left( M_\Delta, 1/\left(1+z\right), \Omega_m, \Omega_b, \sigma_8, h_0\right)$.
As the intrinsic scatter is constant (within few percents) we didn't further parametrise it and assumed it to be independent of mass, redshift and cosmology.
The functional form of the fit parameters in Eq. \ref{eq:mcz}, with a dependency on cosmology is as  follows:

\begin{equation}
\begin{split}
{\rm ln}A &=& A_0 + \alpha_m \Log \left(\frac{\Omega_m}{\Omega_{m,p}}\right) +  \alpha_b \Log \left(\frac{\Omega_b}{\Omega_{b,p}}\right)  +\\
&& + \alpha_\sigma \Log \left(\frac{\sigma_8}{\sigma_{8,p}}\right)  + \alpha_h \Log \left(\frac{h_0}{h_{0,p}}\right)\\
{\rm ln}B &=& B_0 + \beta_m \Log \left(\frac{\Omega_m}{\Omega_{m,p}}\right) +  \beta_b \Log \left(\frac{\Omega_b}{\Omega_{b,p}}\right)  +\\
&& + \beta_\sigma \Log \left(\frac{\sigma_8}{\sigma_{8,p}}\right)  + \beta_h \Log \left(\frac{h_0}{h_{0,p}}\right)\\
{\rm ln}C &=& C_0 + \gamma_m \Log \left(\frac{\Omega_m}{\Omega_{m,p}}\right) +  \gamma_b \Log \left(\frac{\Omega_b}{\Omega_{b,p}}\right)  +\\
&& + \gamma_\sigma \Log \left(\frac{\sigma_8}{\sigma_{8,p}}\right)  + \gamma_h \Log \left(\frac{h_0}{h_{0,p}}\right)
\end{split}
\label{eq:abc}
\end{equation}

\begin{table*}
    \caption{Pivots and best fit parameters for the cosmology-redshift-mass-concentration plane and its dependency on cosmology as in Eq. \ref{eq:mcz} and Eq. \ref{eq:abc} for concentration overdensities of $\Delta=\Delta_{\vir},\Delta_{200c},\Delta_{500c},\Delta_{2500c}$ and $\Delta_{200m}.$     The pivots  $\Omega_{m,p},\Omega_{b,p},\sigma_8$ and $h_0$ in Eq. \ref{eq:abc} are the cosmological parameters of C8 as in Table \ref{tbl:sims} ($\Omega_m=0.272, \Omega_b=0.0456, \sigma_8=0.809, h_0=0.704$). Pivots $a_p$ and $M_p$ are respectively median of scale factor an mass of all haloes. Errors on $A_0,B_0,C_0$ and $\sigma$ are omitted as they are all $<0.001\%.$ The package \texttt{hydro\_mc} contains a script that utilises this relation (\linka{http://github.com/aragagnin/hydro_mc/blob/master/examples/sample_mc.py}).}
    %\begin{tabular}{|l|D{.}{x}{10} llll}
    %\begin{tabular}{|l|*1{d{0.0}} llll}
    \begin{tabular}{rrrrrr}
\hline
  \multirow{1}{*}{Param}  &\multicolumn{5}{c}{Overdensity}\\
                                 & vir                     & 200c                    & 500c                    & 2500c                   & 200m                    \\
\hline
 $M_p [M_{\odot}]$   & $19.9\times10^{13}$  & $17.4\times10^{13}$  & $13.7\times10^{13}$  & $6.9\times10^{13}$   & $22.4\times10^{13}$  \\
 $a_p$           & $0.877$             & $0.877$             & $0.877$             & $0.877$             & $0.877$             \\
 $A_0$           & $1.50$              & $1.24$              & $0.86$              & $0.13$              & $1.69$              \\
 $B_0$           & $-0.04$             & $-0.05$             & $-0.05$             & $-0.03$             & $-0.04$             \\
 $C_0$           & $0.52$              & $0.20$              & $0.19$              & $0.11$              & $0.91$              \\
 $\alpha_m$      & $0.454 \pm 0.041 $  & $0.632 \pm 0.042 $  & $0.662 \pm 0.042 $  & $0.759 \pm 0.055 $  & $0.227 \pm 0.037 $  \\
 $\alpha_b$      & $-0.249 \pm 0.040 $ & $-0.246 \pm 0.038 $ & $-0.235 \pm 0.049 $ & $-0.272 \pm 0.134 $ & $-0.266 \pm 0.035 $ \\
 $\alpha_\sigma$ & $0.554 \pm 0.030 $  & $0.561 \pm 0.034 $  & $0.519 \pm 0.047 $  & $0.422 \pm 0.050 $  & $0.528 \pm 0.022 $  \\
 $\alpha_h$      & $-0.005 \pm 0.030 $ & $-0.026 \pm 0.016 $ & $-0.031 \pm 0.065 $ & $-0.021 \pm 0.167 $ & $0.016 \pm 0.028 $  \\
 $\beta_m$       & $-0.122 \pm 0.001 $ & $-0.118 \pm 0.001 $ & $-0.112 \pm 0.001 $ & $-0.116 \pm 0.001 $ & $-0.116 \pm 0.001 $ \\
 $\beta_b$       & $0.117 \pm 0.005 $  & $0.112 \pm 0.004 $  & $0.126 \pm 0.005 $  & $0.289 \pm 0.007 $  & $0.115 \pm 0.008 $  \\
 $\beta_\sigma$  & $0.051 \pm 0.003 $  & $0.056 \pm 0.002 $  & $0.088 \pm 0.004 $  & $0.103 \pm 0.005 $  & $0.050 \pm 0.006 $  \\
 $\beta_h$       & $-0.079 \pm 0.013 $ & $-0.044 \pm 0.009 $ & $-0.156 \pm 0.014 $ & $-0.342 \pm 0.017 $ & $-0.094 \pm 0.027 $ \\
 $\gamma_m$      & $0.240 \pm 0.006 $  & $0.352 \pm 0.007 $  & $0.346 \pm 0.009 $  & $0.384 \pm 0.011 $  & $-0.043 \pm 0.009 $ \\
 $\gamma_b$      & $-0.126 \pm 0.034 $ & $-0.039 \pm 0.040 $ & $-0.045 \pm 0.051 $ & $-0.133 \pm 0.062 $ & $-0.063 \pm 0.053 $ \\
 $\gamma_\sigma$ & $0.664 \pm 0.027 $  & $0.767 \pm 0.026 $  & $0.856 \pm 0.032 $  & $0.846 \pm 0.046 $  & $0.635 \pm 0.039 $  \\
 $\gamma_h$      & $-0.030 \pm 0.109 $ & $-0.276 \pm 0.112 $ & $-0.347 \pm 0.136 $ & $0.003 \pm 0.171 $  & $-0.405 \pm 0.135 $ \\
 $\sigma$        & $0.388 \pm 0.001 $  & $0.384 \pm 0.001 $  & $0.377 \pm 0.001 $  & $0.383 \pm 0.001 $  & $0.388 \pm 0.001 $  \\
\hline
\end{tabular}
    \label{tbl:megafit}
\end{table*}

The fit has been performed for $\Delta=\Delta_{\vir}, \Delta_{200c}, \Delta_{500c}, \Delta_{2500c}$ and $\Delta_{200m}$  by maximising the Likelihood as in Eq \ref{eq:l}.
\grassetto{
Table \ref{tbl:megafit} shows the results with cosmological parameter pivots at the reference cosmology C8.
}

\grassetto{
Given the high number of free parameters, in order to not underestimate possible sources of errors in the fit, we decided to evaluated uncertainties as follows in \cite{2019Singh}: (1) we first re-performed the fit for each simulation by  setting its own cosmological parameter as pivot values;  (2) then for each parameter except $A_0, B_0, C_0,$ we considered the standard deviation of the parameter values in the previous fits and set it as uncertainty  in Table \ref{tbl:megafit}; (3) parameters $A_0, B_0, C_0$ are presented without uncertainty because the error obtained from the Hessian matrix is negligible compared to the scatter parameter $\sigma.$
Being this work first necessary step towards a cosmology-dependent mass-concentration relation, these parameters may be constrained with more precision in future simulation campaigns.
}

From the above fit we find that the normalisation ($\alpha$ parameters) is mainly
affected by $\Omega_m$ and $\sigma_8$. 
The slope of the mass-concentration plane  ($\beta$ parameters) has a weak dependency on cosmology.
However, the logarithmic mass slope is pushed towards negative values by an increase in $\Omega_m$ and $h_0$
(i.e. $\beta_m$ and $\beta_h<0$), while it is pushed towards positive values by an increase in $\Omega_b$ and  
$\sigma_8$ (since $\beta_b$ and $\beta_\sigma>0$).
This behaviour is also shown in Figure \ref{fig:obs}. Note that, C1 and C2 have opposite mass-dependency 
with respect to the other runs. Although the trend can be positive for some cosmologies 
(see Table \ref{tbl:megafit} and Figure \ref{fig:mczs}), the slope is always close to zero.
The redshift dependency ($\gamma$ parameters) is driven by both $\sigma_8$ and $\Omega_m,$ while a high baryon 
fraction can lower the dependency (see parameter $\gamma_h$).
The scatter is nearly constant for all  the overdensities with a value close to $0.38$ 
\junof{and even if it is of the same order of the difference between Mc relations of different cosmologies (see shaded area in Fig. \ref{fig:obs}),  in the next sub-section we will show that statistical studies  on  large  samples of galaxy clusters are still   affected by the cosmological dependency of  Mc relations.}

Since the logarithmic slope of the mass has a weak dependency on cosmology, we provide a similar fit as the one in this
section without $B$ having any cosmological dependencies i.e. $B=B_0$ in Appendix \ref{ap:mcc} 
(see Table \ref{tbl:megafit_lite}). 
In Appendix \ref{ap:mcc} (see Table \ref{tbl:megafit_dm_lite}) we also provide the same reduced fit parameters with the 
scale radius computed on the dark matter density profile.

\subsection{Impact on inferred weak-lensing masses}

\begin{figure}
  \caption{Top  panel: surface density of mock data points (gray line), and best fit realisations with DMO Mc relation  priors (red line) and this work (blue line). Middle and shows marginalized posterior parameter distributions of log$_{10} M_{200c}$ and $c_{200}$  with DMO Mc relation (red and pink lines) and this works Mc relation (blue and cyan lines) of a signle cluster realisation. Bottom panel shows the same for a stack of $100$ GCs. }
   \includegraphics[width=\linewidth]{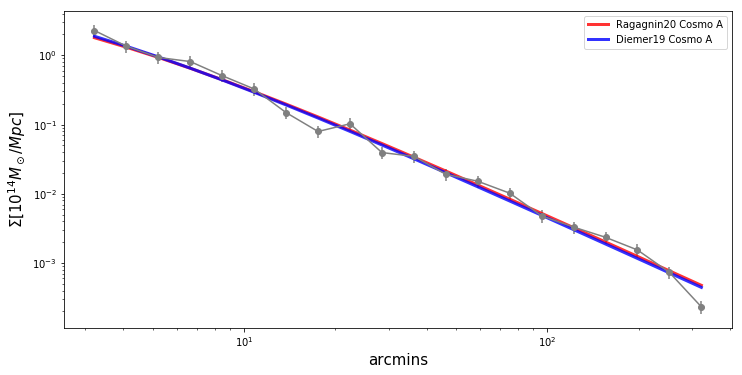}\\
    \includegraphics[width=\linewidth]{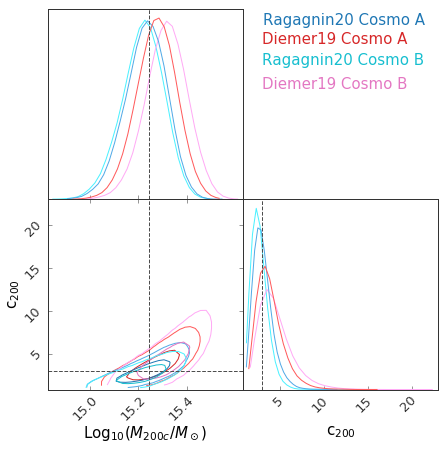}
    \includegraphics[width=\linewidth]{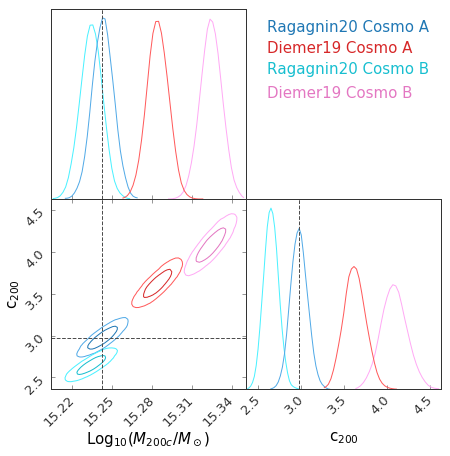}
\label{fig:rdprof_3}
\end{figure}

\juno{There are differences between the Mc relation extracted from our simulations at different cosmologies} \junof{ and the ones from DMO simulations.}
\juno{When Mc relations are used to provide priors and interpret the weak-lensing signal of} \junof{non-ideal NFW} \juno{cluster samples, the different Mc relations will ultimately lead to different inferred masses and therefore different cosmological constraints from cluster number counts experiments. In fact, works as \cite{2017MNRAS.465.3361Henson}  show that it is possible to correctly recover halo masses from mock observations of both DMO and hydro-simulations by using their respective mass-concentration relations. 
On the other hand, in low signal-to-noise conditions, weak-lensing mass calibration typically constrains the total observed mass using Mc relations derived from DMO simulations~\citep[see, e.g., ][]{2017MNRAS.469.4899Melchior}.
In the following, we quantify and discuss this effect on a simplified example.}

\junof{To this end, we create a simulated projected surface density  profile} \junof{of an NFW model of the RXC J2248.7-4431 cluster~\citep{2013MNRAS.432.1455Gruen}, at $z=0.436$ with mass  $M_{200c}= 1.75 \cdot 10^{15}\Msun$~\citep{2015ascl.soft11017Melchior}. The simulated profile is generated using  Eq. 41 in \cite{2001MNRAS.321..155L} with a concentration $c_{200c}$ from our Mc relation (see Table \ref{tbl:megafit}) and with cosmological parameters  $\Omega_m=0.27, \Omega_b=0.05, \sigma_8=0.8, h_0=0.67$. We mimic a simplified observed radial profile sampled with $20$ logarithmic equally spaced radial bins from $3$ to $30$ arcmin. To each data-point we assigned an associated error in order to simulate typical weak-lensing observational conditions ($S/N=5$) of a massive clusters in a photometric survey like the Dark Energy Survey  ~\citep[DES,][]{2015ascl.soft11017Melchior}.
We test a simplified mass calibration process by fitting the above described density profile with a gaussian likelihood for each simulated projected density radial bin $\Sigma_i$:
\begin{equation}
    \mathcal{L} = \prod_i P(\Sigma_i|M_{200c},c_{200c},\Delta_{\Sigma,i}).
\end{equation}
We adopt a flat prior on log$M_{200c}$ and test the impact of adopting the following different priors for the concentration:
}

\begin{itemize}
  \item \juno{Ragagnin20: The Mc relation with a lognormal scatter $\sigma_{\mathrm{ln}c|M} = 0.38$ as presented in this work.}
  \item\juno{Diemer19: The DMO Mc relation proposed in \cite{2019ApJ...871..168Diemer}\footnote{We used the python package \linka{https://bdiemer.bitbucket.io/colossus/} \citep[see][]{2018ApJS..239...35D}.} with a lognormal scatter $\sigma_{\mathrm{ln}c|M} = 0.39$.}%,  with  cosmological parameters as above, $n_s=0.963$ and its scatter $\sigma_{Log_{10} c200c} = 0.17.$ }
\end{itemize}
% 0.26, 0.04, 0.6, .66
\junof{To show the impact on mass-calibration of  Ragagnin20 and the dependency on cosmological parameters, we perform the calibration both at the correct input cosmology (from here on Cosmo A) and with cosmological parameters randomly extracted from the posterior distribution of the cosmological parameters  derived by SPT cluster number counts \citep[][ $\Omega_m=0.26, \Omega_b=0.04, \sigma_8=0.6, h_0=0.66$ from here on Cosmo B]{2019ApJ...878...55B}.}

\junof{Figure \ref{fig:rdprof_3} (top panel) shows the ideal un-perturbed mock profile and best fit realisations of NFW profiles produced using Ragagnin20 (red line) and Diemer19~(blue line). The Mc relation presented in this work has a lower concentration normalisation than Diemer19 (Appendix \ref{ap:effects}), and thus Ragagnin20 produce lower values of surface densities near the centre and higher values on the outskirts with respect to DMO mass-concentration relations.
Different prior assumptions on the Mc relation affect the inferred mass, as we see in Figure \ref{fig:rdprof_3} (middle panel)\footnote{Parameter space is sampled with \texttt{emcee}.}. While the posterior derived assuming the Mc relations of Ragagnin20 and of Diemer19 are in good agreement, the best fit mass recovered with Diemer19 is $\approx10\%$ higher compared to the one derived with Ragagnin20. This can be better appreciated in the bottom panel of Figure \ref{fig:rdprof_3}, where we instead simulate the mass calibration of a stack of $100$ clusters \citep{2017MNRAS.469.4899Melchior}.
We therefore mimic a $S/N=50$ stacked average profile and decrease by a factor of $\sqrt{100}$ the intrinsic scatter around the Mc relation in the prior. Assuming the wrong Cosmo B cosmology, we would recover biased results using both the Ragagnin20 and the Diemer19 Mc relation, even if the marginalized posterior on the mass would be almost unbiased for the Ragagnin20 analysis.
Furthermore we note that fixing Cosmo B cosmolgy instead of the correct input Cosmo A cosmolgy would result in a slightly smaller mass for Ragagnin20 and in a slightly larger mass for Diemer19. While a more sophisticated analysis including a treatment of systematic uncertainties and a  self-consistent exploration of the cosmological parameters  is beyond the purpose of this work, this simple exercise highlights the importance of a correct modelization of the cosmological dependence of the Mc relation in the the weak-lensing analysis of cluster samples for cosmological purposes.  
}

\juno{
We stress that in this experiment we wanted  to mimic the procedure of most observational works, thus we didn't model baryon component of DMO simulations.
}

%%%%%%%%%%%%%%%%%%%%%%%%%%%%%%%%%%%%%%%%%%%%%%%%%%%%%%
%%%%%%%%%%%%%      Halo masses      %%%%%%%%%%%%%%%%%%
%%%%%%%%%%%%%%%%%%%%%%%%%%%%%%%%%%%%%%%%%%%%%%%%%%%%%%

\section{Halo masses conversion}
\label{sec:mmac}
%In the following subsections we study the possibility of converting masses from one overdensity  to the other (e.g. the problem of obtaining $M_{200}$ given $M_{500}$).
In the following subsections, we present and compare different methods of converting masses between  overdensities.
We also provide a direct fit for converting  masses 
(i.e. SUBFIND masses) from $\Delta1$ to $\Delta2$ (thus without using the Mc relation), in order to study the origin of the scatter coming from the conversions. 
This kind of conversions is used in computing the sparsity of haloes (i.e. ratio of masses in two overdensities), which itself can probe cosmological parameters \citep{2018ApJ...862...40C,2019MNRAS.487.4382C} and  dark energy models \citep{2014MNRAS.437.2328B}.

\subsection{Mass-mass conversion using Mc relation}
\label{seq:mmconvmc}
In this section, we tackle the problem of converting masses via a Mc relation.
By combining the definition of mass $M_\Delta$ (see Eq. \ref{eq:mdelta}) and the fact that the matter profile only depends on a proportional parameter $\rho_0$ and a scale radius $r_s,$ we get

\begin{equation}
M_\Delta = 4\pi \rho_0 r_s^3 f(c_\Delta) = \Delta\frac{4}{3}\pi R_\Delta^3  \rho_c,
\label{eq:m_delta}
\end{equation}

For a NFW profile as in Eq. \ref{eq:nfw},

\begin{equation}
\begin{split}
f(c_\Delta) &=& \ln(1+c_\Delta)-\frac{c_\Delta}{1+c_\Delta}.
\end{split}
\label{eq:fc}
\end{equation}

Combining Eq. \ref{eq:m_delta} and \ref{eq:fc} gives the following mass conversion formula:

\begin{equation}
\begin{split}
\begin{dcases}
M_{\Delta2} =  M_{\Delta1} \left(\frac{c_{\Delta2}}{c_{\Delta1}}\right)^3\frac{\Delta_2}{\Delta_1}\\
\\
c_{\Delta,2} = c_{\Delta,1} \cdot \left(\frac{\Delta_1}{\Delta_2} \frac{f(c_{\Delta,2})}{f(c_{\Delta,1})}\right)^{\frac{1}{3}}.
\end{dcases}
\end{split}
\label{eq:mm}
\end{equation}

From the second part of Eq. \ref{eq:mm} it is possible to  evaluate the concentration  $c_{\Delta2}$ as a function of only $c_{\Delta1}$   \citep[as in Appendix C of][]{2003ApJ...584..702H}.

 Eq. \ref{eq:mm} can be used to estimate the theoretical scatter $\sigma_{\mathrm theo}$ obtained in the mass conversion by analytically propagating the uncertainties of the mass-concentration relation, namely:

\begin{equation}
    \sigma_{\mathrm theo} = \frac{1}{{M_\dd}}\frac{dM_\dd}{dc_\du} \sigma_{c,\du},
    \label{eq:sigmat}
\end{equation}

 where $M_\dd$ is the converted mass, $c_\du$ the concentration in the original overdensity $\Delta_1$ and
 $\sigma_{c,\du}$ is the uncertainty in the concentration (in our case it is the scatter in the Mc relation).
Appendix \ref{ap:scat} describes how to  obtain the theoretical scatter one would expect given a perfectly NFW profile.

%One would expect that  the mass-mass conversion  derived by a mass-concentration relation,  has  several sources of error:
There are several sources of error in the mass-mass conversion  derived by a mass-concentration relation:
(i)  the intrinsic  scatter of the Mc relation,
(ii) the fact that profiles are not perfectly NFW and thus Eq. \ref{eq:fc} is not the best choice for this conversion; (iii) the cosmology-redshift-mass-concentration fit (as in Table \ref{tbl:megafit}) may not be optimal.

To further study the sources of uncertainties in this conversion, we fit SUBFIND halo masses between two overdensities\footnote{The package \texttt{hydro\_mc} contains a sample script to convert masses between two overdensities by using the mass-concentration relation presented in this paper (\linka{http://github.com/aragagnin/hydro_mc/blob/master/examples/sample_mm_from_mc_relation.py}).}, and compare the two conversion methods.

\subsection{$M_{\Delta1}$-$M_{\Delta2}$ (M-M) plane}
\label{seq:mmconv}

In this subsection we perform a direct fit between halo masses (i.e. SUBFIND masses), as a function of redshift and cosmological parameter.
The reason of this fit is twofold: (1) we want to study the uncertainty introduced in the conversion of the previous subsection and (2) we want to provide a way of converting masses without any assumption on their concentration and NFW density profile.

For each pair of overdensities we performed a fit of the mass $M_{\Delta2} \left( M_{\Delta1}, 1/\left(1+z\right), \Omega_m, \Omega_b, \sigma_8, h_0\right)$ with the following functional form:
\begin{equation}
\Log\ M_{\Delta2} \left(M_{\Delta1}, a\right) = \Log A + B \Log \left(\frac{M_{\Delta1}}{M_p}\right) + C \Log \left(\frac{a}{a_p}\right)
\label{eq:mmz}
\end{equation}
where $A,B,C$ parameters are parametrised with cosmology as in Eq.  \ref{eq:abc}.

%
% tbl:megafit_mm
%
\begin{table*}
    \caption{Fit parameters for  Eq. \ref{eq:mmz} and Eq. \ref{eq:abc} for  overdensities   $\Delta_{\texttt vir},\Delta_{200c},\Delta_{500c}, \Delta_{2500c}$ and $\Delta_{200m}$.  Pivots are as in Table \ref{tbl:megafit}.  Errors on $A_0,B_0,C_0$ and $\sigma$ are omitted as they are all $<0.001\%.$  The package \texttt{hydro\_mc} contains a script that utilises this relation (\linka{http://github.com/aragagnin/hydro_mc/blob/master/examples/sample_mm.py}).}
    \begin{tabular}{r rrr rrr}
\hline
   \multirow{2}{*}{Param}  &\multicolumn{6}{c}{From overdensity $\rightarrow$ to overdensity }\\
                 & $\texttt vir \rightarrow 200c$            & $\texttt vir \rightarrow 500c$            & $\texttt vir \rightarrow 2500c$            & $200c \rightarrow \texttt vir $            & $200c \rightarrow 500c$           & $200c \rightarrow 2500c$           \\
\hline
 $M_p [M_{\odot}]$   & $19.9\times10^{13}$  & $19.9\times10^{13}$  & $19.9\times10^{13}$  & $17.4\times10^{13}$  & $17.4\times10^{13}$  & $17.4\times10^{13}$  \\
 $a_p$           & $0.877$             & $0.877$             & $0.877$             & $0.877$             & $0.877$             & $0.877$             \\
 $A_0$           & $32.72$             & $32.31$             & $31.34$             & $32.99$             & $32.39$             & $31.41$             \\
 $B_0$           & $1.00$              & $1.00$              & $0.93$              & $0.99$              & $0.99$              & $0.92$              \\
 $C_0$           & $-0.24$             & $-0.24$             & $-0.01$             & $0.23$              & $0.00$              & $0.22$              \\
 $\alpha_m$      & $0.165 \pm 0.007 $  & $0.295 \pm 0.015 $  & $0.619 \pm 0.026 $  & $-0.156 \pm 0.006 $ & $0.125 \pm 0.009 $  & $0.461 \pm 0.021 $  \\
 $\alpha_b$      & $0.003 \pm 0.011 $  & $-0.020 \pm 0.033 $ & $-0.162 \pm 0.069 $ & $-0.003 \pm 0.009 $ & $-0.020 \pm 0.023 $ & $-0.150 \pm 0.062 $ \\
 $\alpha_\sigma$ & $0.048 \pm 0.008 $  & $0.177 \pm 0.016 $  & $0.574 \pm 0.016 $  & $-0.035 \pm 0.006 $ & $0.120 \pm 0.009 $  & $0.539 \pm 0.016 $  \\
 $\alpha_h$      & $-0.045 \pm 0.004 $ & $-0.105 \pm 0.013 $ & $-0.054 \pm 0.068 $ & $0.037 \pm 0.005 $  & $-0.060 \pm 0.012 $ & $-0.042 \pm 0.065 $ \\
 $\beta_m$       & $-0.016 \pm 0.001 $ & $-0.043 \pm 0.001 $ & $-0.076 \pm 0.001 $ & $0.015 \pm 0.001 $  & $-0.025 \pm 0.001 $ & $-0.058 \pm 0.001 $ \\
 $\beta_b$       & $0.030 \pm 0.001 $  & $0.096 \pm 0.003 $  & $0.203 \pm 0.004 $  & $-0.026 \pm 0.001 $ & $0.062 \pm 0.003 $  & $0.171 \pm 0.003 $  \\
 $\beta_\sigma$  & $-0.024 \pm 0.001 $ & $-0.043 \pm 0.002 $ & $-0.010 \pm 0.003 $ & $0.017 \pm 0.001 $  & $-0.020 \pm 0.002 $ & $0.006 \pm 0.002 $  \\
 $\beta_h$       & $-0.007 \pm 0.003 $ & $-0.038 \pm 0.008 $ & $-0.203 \pm 0.009 $ & $0.015 \pm 0.003 $  & $-0.025 \pm 0.009 $ & $-0.180 \pm 0.009 $ \\
 $\gamma_m$      & $0.159 \pm 0.001 $  & $0.213 \pm 0.002 $  & $0.379 \pm 0.001 $  & $-0.153 \pm 0.002 $ & $0.052 \pm 0.003 $  & $0.211 \pm 0.002 $  \\
 $\gamma_b$      & $-0.050 \pm 0.005 $ & $-0.115 \pm 0.010 $ & $-0.232 \pm 0.006 $ & $0.057 \pm 0.009 $  & $-0.052 \pm 0.016 $ & $-0.080 \pm 0.013 $ \\
 $\gamma_\sigma$ & $0.123 \pm 0.003 $  & $0.354 \pm 0.009 $  & $0.555 \pm 0.009 $  & $-0.107 \pm 0.005 $ & $0.244 \pm 0.007 $  & $0.486 \pm 0.005 $  \\
 $\gamma_h$      & $0.036 \pm 0.011 $  & $0.026 \pm 0.034 $  & $0.039 \pm 0.035 $  & $-0.035 \pm 0.019 $ & $-0.038 \pm 0.029 $ & $-0.175 \pm 0.031 $ \\
 $\sigma$        & $0.065 \pm 0.001 $  & $0.158 \pm 0.001 $  & $-0.312 \pm 0.001 $ & $0.056 \pm 0.001 $  & $0.113 \pm 0.001 $  & $-0.296 \pm 0.001 $ \\
\hline
\end{tabular}
    \begin{tabular}{r rrr rrr}
\hline
   \multirow{2}{*}{Param}    &\multicolumn{6}{c}{From overdensity $\rightarrow$ to overdensity }\\
                 & $500c \rightarrow \texttt vir$            & $500c \rightarrow 200c$           & $500c \rightarrow 2500c$          & $2500c \rightarrow  \texttt vir$           & $2500c \rightarrow 200c$          & $200c \rightarrow 500c$          \\
                 & $500c - vir$        & $500c - 200c$       & $500c - 2500c$      & $2500c - vir$       & $2500c - 200c$      & $2500c - 500c$      \\
\hline
 $M_p [M_{\odot}]$   & $13.7\times10^{13}$  & $13.7\times10^{13}$  & $13.7\times10^{13}$  & $6.9\times10^{13}$   & $6.9\times10^{13}$   & $6.9\times10^{13}$   \\
 $a_p$           & $0.877$             & $0.877$             & $0.877$             & $0.877$             & $0.877$             & $0.877$             \\
 $A_0$           & $33.12$             & $32.93$             & $31.58$             & $33.33$             & $33.14$             & $32.77$             \\
 $B_0$           & $0.99$              & $1.00$              & $0.93$              & $1.02$              & $1.03$              & $1.03$              \\
 $C_0$           & $0.25$              & $0.02$              & $0.22$              & $0.16$              & $-0.08$             & $-0.10$             \\
 $\alpha_m$      & $-0.264 \pm 0.013 $ & $-0.114 \pm 0.007 $ & $0.335 \pm 0.012 $  & $-0.563 \pm 0.030 $ & $-0.414 \pm 0.023 $ & $-0.307 \pm 0.014 $ \\
 $\alpha_b$      & $0.003 \pm 0.031 $  & $0.006 \pm 0.021 $  & $-0.125 \pm 0.045 $ & $0.093 \pm 0.142 $  & $0.100 \pm 0.126 $  & $0.090 \pm 0.091 $  \\
 $\alpha_\sigma$ & $-0.111 \pm 0.015 $ & $-0.088 \pm 0.009 $ & $0.409 \pm 0.016 $  & $-0.342 \pm 0.009 $ & $-0.316 \pm 0.017 $ & $-0.255 \pm 0.022 $ \\
 $\alpha_h$      & $0.084 \pm 0.025 $  & $0.058 \pm 0.019 $  & $0.015 \pm 0.080 $  & $0.103 \pm 0.195 $  & $0.066 \pm 0.184 $  & $0.019 \pm 0.155 $  \\
 $\beta_m$       & $0.034 \pm 0.001 $  & $0.019 \pm 0.001 $  & $-0.033 \pm 0.001 $ & $0.063 \pm 0.001 $  & $0.049 \pm 0.001 $  & $0.029 \pm 0.001 $  \\
 $\beta_b$       & $-0.083 \pm 0.005 $ & $-0.053 \pm 0.002 $ & $0.115 \pm 0.002 $  & $-0.300 \pm 0.001 $ & $-0.264 \pm 0.003 $ & $-0.189 \pm 0.003 $ \\
 $\beta_\sigma$  & $0.033 \pm 0.003 $  & $0.019 \pm 0.001 $  & $0.031 \pm 0.001 $  & $-0.019 \pm 0.001 $ & $-0.035 \pm 0.002 $ & $-0.045 \pm 0.001 $ \\
 $\beta_h$       & $0.064 \pm 0.012 $  & $0.036 \pm 0.005 $  & $-0.197 \pm 0.005 $ & $0.412 \pm 0.003 $  & $0.382 \pm 0.007 $  & $0.320 \pm 0.006 $  \\
 $\gamma_m$      & $-0.190 \pm 0.002 $ & $-0.033 \pm 0.002 $ & $0.162 \pm 0.001 $  & $-0.306 \pm 0.001 $ & $-0.159 \pm 0.001 $ & $-0.134 \pm 0.003 $ \\
 $\gamma_b$      & $0.101 \pm 0.009 $  & $0.005 \pm 0.012 $  & $-0.081 \pm 0.005 $ & $0.152 \pm 0.005 $  & $0.075 \pm 0.010 $  & $0.047 \pm 0.018 $  \\
 $\gamma_\sigma$ & $-0.373 \pm 0.006 $ & $-0.281 \pm 0.008 $ & $0.223 \pm 0.003 $  & $-0.638 \pm 0.008 $ & $-0.532 \pm 0.014 $ & $-0.268 \pm 0.010 $ \\
 $\gamma_h$      & $-0.017 \pm 0.023 $ & $0.130 \pm 0.033 $  & $-0.083 \pm 0.012 $ & $0.037 \pm 0.024 $  & $0.147 \pm 0.038 $  & $0.043 \pm 0.042 $  \\
 $\sigma$        & $0.129 \pm 0.001 $  & $0.096 \pm 0.001 $  & $-0.235 \pm 0.001 $ & $0.242 \pm 0.001 $  & $0.228 \pm 0.001 $  & $0.182 \pm 0.001 $  \\
\hline
\end{tabular}
    \label{tbl:megafit_mm}
\end{table*}

%
% tbl:megafit_mm200m
%
\begin{table*}
    \caption{Fit parameters for  Eq. \ref{eq:mmz} and Eq. \ref{eq:abc} between overdensities of  $\Delta_{200c}$ to $\Delta_{200m}.$  Errors on $A_0,B_0,C_0$ and $\sigma$ are omitted as they are all $<0.001\%.$  Pivots are as in Table \ref{tbl:megafit}. The package \texttt{hydro\_mc} contains a script that utilises this relation (\linka{http://github.com/aragagnin/hydro_mc/blob/master/examples/sample_mm.py}).}
    \begin{tabular}{r rr}
\hline
   \multirow{2}{*}{Param}     &\multicolumn{2}{c}{From overdensity $\rightarrow$ to overdensity }\\
                 & $200c \rightarrow 200m$           & $200m \rightarrow 200c$            \\
\hline
 $M_p [M_{\odot}]$   & $17.4\times10^{13}$  & $22.4\times10^{13}$  \\
 $a_p$           & $0.877$             & $0.877$             \\
 $A_0$           & $33.11$             & $32.71$             \\
 $B_0$           & $0.99$              & $1.01$              \\
 $C_0$           & $0.46$              & $-0.49$             \\
 $\alpha_m$      & $-0.288 \pm 0.013 $ & $0.318 \pm 0.016 $  \\
 $\alpha_b$      & $-0.017 \pm 0.015 $ & $0.014 \pm 0.019 $  \\
 $\alpha_\sigma$ & $-0.053 \pm 0.011 $ & $0.088 \pm 0.019 $  \\
 $\alpha_h$      & $0.078 \pm 0.009 $  & $-0.103 \pm 0.003 $ \\
 $\beta_m$       & $0.031 \pm 0.001 $  & $-0.035 \pm 0.001 $ \\
 $\beta_b$       & $-0.040 \pm 0.003 $ & $0.063 \pm 0.003 $  \\
 $\beta_\sigma$  & $0.029 \pm 0.002 $  & $-0.050 \pm 0.002 $ \\
 $\beta_h$       & $0.017 \pm 0.006 $  & $-0.011 \pm 0.007 $ \\
 $\gamma_m$      & $-0.313 \pm 0.002 $ & $0.358 \pm 0.002 $  \\
 $\gamma_b$      & $0.062 \pm 0.008 $  & $-0.102 \pm 0.010 $ \\
 $\gamma_\sigma$ & $-0.201 \pm 0.007 $ & $0.246 \pm 0.010 $  \\
 $\gamma_h$      & $0.026 \pm 0.022 $  & $0.033 \pm 0.036 $  \\
 $\sigma$        & $0.084 \pm 0.001 $  & $-0.102 \pm 0.001 $ \\
\hline
\end{tabular}
    \label{tbl:megafit_mm200m}
\end{table*}

Table \ref{tbl:megafit_mm} show the results of the mass-mass conversion fit between critical overdensities, while Table \ref{tbl:megafit_mm200m} show the conversion fit parameters between $\Delta_{200c}$ and $\Delta_{200m}$.
The conversion relation  has a  strong dependency on $\sigma_8$ and a weak dependency on $h_0$ 
(see $\alpha_m,\beta_m,\gamma_m$ parameters).
%We quantitatively discuss these results in the discussion Section \ref{sec:conclu}.

\subsection{Uncertainties in mass conversions}

When converting between masses at different overdensities,  we are interested in the the following sources of uncertainty:
 \begin{itemize}
\item $\sigma_{M-M(Mc)}:$  the scatter  from the mass-mass conversion obtained  with the aid of our Mc relation found in Sec. \ref{seq:mmconvmc}.
\item  $\sigma_{M-M(c)}:$  the scatter obtained from a  conversion between the true values of $M_{\Delta1}$ and $c_{\Delta1}$ of a given halo to  $M_{\Delta2}$ (i.e. using only Eq. \ref{eq:mm}). We use this scatter in order to estimate the error coming from non-NFWness (i.e. deviation from perfect NFW density profile).
\item  $ \sigma_{\textrm theo}:$ the scatter obtained by analytically propagating the Mc log-scatter ($\approx 0.38$ as in Table \ref{tbl:megafit}) with Eq.  \ref{eq:sigmat}. This value estimates the uncertainty coming from the intrinsic scatter of the Mc relation alone.
\item $\tilde \sigma_{Mc}:$ the scatter that is supposed to be introduced by a non-ideal cosmology-redshift-mass-concentration fitting formula.
\item  $\sigma_{M-M}:$ the intrinsic scatter of M-M conversion using Table \ref{tbl:megafit_mm} presented in Sec. \ref{seq:mmconv}.
\end{itemize}

In a simplistic approach,  the quadrature sum of the scatter coming from non-NFWness ( $\sigma_{M-M(c)}$), the theoretical scatter  ($ \sigma_{\texttt theo}$) and the scatter due to a non-ideal Mc fit ($\tilde \sigma_{Mc}$),  should all add up to the scatter in the mass-mass conversion using a mass-concentration relation: 

\begin{equation}
   \sigma_{M-M(Mc)}^2 =  \sigma_{M-M(c)}^2 +  \sigma_{\texttt theo}^2 + \tilde \sigma_{Mc}^2.
   \label{eq:scatteri}
\end{equation}

\subsection{Obtaining $M_{200c}$ from  $M_{500c}$ or  $M_{2500c}$}

In this subsection we test mass conversion to  $M_{200c}$ given $M_{500c}$ or $M_{500c}.$ We compare results obtained using the technique described in Sec \ref{seq:mmconvmc} 
against the mass-mass relation from Eq. \ref{eq:mmz}.

We tested the conversion $M_{500c}\rightarrow M_{200c}$ in the mass regime of $M_{200c}\approx 10^{14}-10^{15}\Msun$ and found  the following  scatter values: $\sigma_{M-M(Mc)} =  0.09,$  
$\sigma_{M-M(c)} =  0.04,$ and $\sigma_{\texttt theo} = 0.07$  
by converting masses using Sec. \ref{seq:mmconvmc} and $\sigma_{M-M} = 0.07$ by   using conversion table in Sec.   \ref{seq:mmconv}. 
We are confident to have all uncertainty sources under control because the quadrature sum of all scatters from conversion described in Sec. \ref{seq:mmconv} (i.e. $\sqrt{ \sigma_{M-M(c)}^2 + \sigma_{\texttt theo}^2 } = 0.09$) equals  $\sigma_{M-M(Mc)}$ from Sec.  \ref{seq:mmconvmc} as in Eq. \ref{eq:scatteri}.

We repeat the same conversion for $M_{2500c}\rightarrow M_{200c}$ and find the following scatter values: $\sigma_{M-M(Mc)} = 0.29,$ $\sigma_{M-M(c)} =  0.07,$ $\sigma_{\texttt theo} = 0.24$ by converting masses using Sec. \ref{seq:mmconvmc} and $\sigma_{M-M} = 0.22,$  by   using conversion table in Sec.   \ref{seq:mmconv}. 
%The main difference with the previous  conversion is that in this case the value  $ \sigma_{\texttt theo}$ has a larger value than $\sigma_{M-M}$ fit.
%As a consequence, 
In this conversion, the quadrature sum of the theoretical scatters in Eq. \ref{eq:scatteri} holds only if we attribute an additional source of the uncertainty to a non-ideal  $M_{2500c}$-$c_{2500c}$ relation fit $\tilde \sigma_{Mc}=0.14.$

This means that  a direct mass-mass fit is more precise than a conversion that passes through an Mc relation when converting $M_{2500c}\rightarrow M_{200c}.$

It is interesting  to see that in both scenarios the conversion with the lowest scatter is the one performed with the exact knowledge of both mass and concentration (i.e. $\sigma_{M-M(c)}$ is the lowest).
On the other hand, in the scenario where one only knows the mass of a halo, then the conversion with the lowest uncertainty is the one that uses  relation in Sec. \ref{seq:mmconv} (i.e. with a scatter  $\sigma_{M-M}$).

\section{Discussion on cosmology dependency of masses and concentrations}
\label{sec:discu}

\begin{figure}
  \caption{Change  in concentration parameter (top panel) and sparsity (bottom panel) by fractional changes of cosmological parameters with respect to reference cosmology C8. Masses and sparsities are computing using our fit  in Table \ref{tbl:megafit}  and Table \ref{tbl:megafit_mm} for a halo of $M_{\Delta}=10^{14}\Msun$. }
  \includegraphics[width=\linewidth]{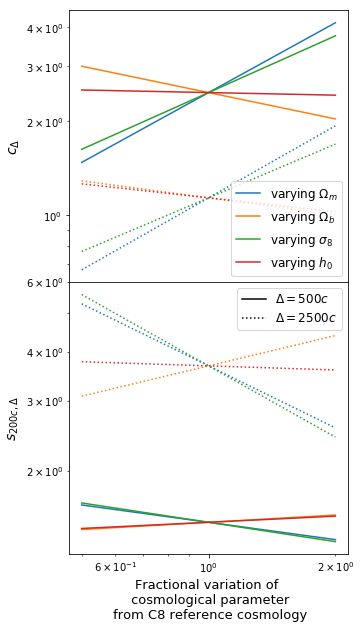}
\label{fig:plot_tables}
\end{figure}

\grassetto{
\junof{The concentration of haloes at fixed mass is a non trivial function of cosmological parameters. We summarise in Figure \ref{fig:plot_tables}~(upper panel) the  variation of $c_{500c}$ as a function of cosmological parameters  for a halo of mass $M_{\Delta}=10^{14}\Msun$ to provide a more intuitive representation.
In general concentration normalisation decreases with baryon fraction $\Omega_b.$
While a small ($\approx2\%$) decrease is expected also on DMO models as \cite{2019ApJ...871..168Diemer},
our change in this mass range is likely associated with feedback from AGN, as an increase of gas fraction implies more energy released by feedback processes, which is known to lower concentration.
The logarithmic mass slope of the Mc (see Table \ref{tbl:megafit}) increases   with $\Omega_b$, in agreement with SNe  feedback being less relevant in massive haloes.
For concentration at $\Delta=2500c$ the situation is less clear. $\Delta=2500c$ is closer to the centre of the halo and depends more strongly on physical processes which are not solely regulated by gravity. The qualitative behaviour is nevertheless consistent with $\Delta=500c$, with the strongest (positive) cosmological dependence on $\sigma_8$ and $\Omega_m$, a weak (negative) dependency on $\Omega_b$, and a weaker one on $h_0$. }  
It is not possible to infer the effect of $\Omega_b$ on the redshift log-slope as its value is mainly driven by $\sigma_8.$ 
}

\grassetto{C1 and C2 simulations show a  positive correlation between mass and concentration.
This is in agreement with \cite{2012MNRAS.423.3018P}, where they found that haloes with low  r.m.s. fluctuation amplitude   $\sigma$ have a concentration that increases with mass. In fact C1 and C2 have extremely low values of $\sigma_8$ (i.e. $\sigma_8<0.7$) which leads to low   r.m.s. fluctuation amplitudes.
}

\grassetto{
When converting masses from higher overdensities to lower overdensities the scatter increases as the difference between overdensities increases (see Table  \ref{tbl:megafit_mm}).
Figure \ref{fig:plot_tables}~(lower panel) shows the variation of   sparsity normalisation   as a function of cosmological parameters. 
The log-slope of the mass dependency ($\beta$ parameters) has almost no dependency on cosmology.
One exception is made by  $s_{200c,2500c},$ where normalisation does depend on $\Omega_b.$
Note that this relation doesn't assume any density profile, thus this dependency cannot be caused by a bad NFW fit. 
This effect is probably due to baryon feedback that at this scale is capable of influencing the total matter density profile.
}

\grassetto{
The redshift dependency ($\gamma$ parameters) is mostly influenced by $\Omega_m$ and $\sigma_8,$ with a contribution that increases with separation between overdensities, which may indicates a different growth of the internal and external regions of the halo.
}
%
% CONCLU
%

\section{Conclusions}
\label{sec:conclu}

\juno{We provided mass-concentration relations and mass conversion relations between overdensities that include dependencies on the cosmological parameters without modelling dynamical state of the simulated haloes.
We showed that mass calibrated with DMO Mc relation can be $\approx 10\%$  higher compared with masses calibrated with our Mc relation. Additionally, cluster-cosmology oriented studies  will benefit from this work since this relation averages over all different dynamical states and includes the average effect of baryon physics.}

For these reasons we performed the following studies:
\begin{itemize}
    \item We provided the fitting functions for the cosmology-redshift-mass-concentration relation in the context of Magneticum hydrodynamic simulations for the critical overdensities $\Delta_\texttt{vir}, \Delta_{200c}, \Delta_{500c}, \Delta_{2500c}$ and mean overdensity $\Delta_{200m})$  (see Sec. \ref{sec:mcac} and Table \ref{tbl:megafit}).
 %\item \juno{We performed some mock mass calibration tests and showed that our Mc relation calibrates different mass values compared to Mc relations coming from DMO simulations.}
    %\item \juno{We found that Magneticum simulations can be used to calibrate   masses of mock profiles obtained from \cite{2010PASJ...62..811O} Mc relation with a scatter of ($\approx4\%$), as opposed to calibration performed with DMO simulation relations that has an average offset of $\approx20\%$.}
    %Mc relation with a precision down to $3\%,$  as opposed to relations of DMO simulations that are on average $\approx20\%$ higher.   We stress the DMO relation from \cite{2015ApJ...799..108D} had no corrections from the baryon components (which may have improved their results) as to mimic mass-calibrations performed in the literature.}
 \item  We explored the possibility of converting masses between overdensities with and without the aid of our mass-concentration relation and, for the latter, we studied the origin of its uncertainty as being caused by (i)  non-NFWness of profiles  and (ii) a non-ideal mass-concentration fit. In particular, when converting masses via an Mc relation, non-NFWness of density profiles accounts for approx. $6\%$ of the scatter. Additionally the conversion between $M_{2500}$ to $M_{200}$ has an additional fractional scatter of $\approx0.15$  caused by the non-ideal mass-concentration relation fit.
\item  In Sec. \ref{sec:discu} we discuss the dependency of halo masses and concentration as a function of cosmological parameters. Although concentration is mainly driven by $\Omega_m$ and $\sigma_8,$ we found that $\Omega_b$ do decreases concentration and  a higher $h_0$ lowers the concentration of the internal part of the halo, probably  because of the related scale-dependent baryon feedback. We also found that the positive mass-concentration trend in $C1$ and $C2$ is due to their low $\sigma_8.$
\end{itemize}

\grassetto{
 We released the python package \texttt{hydro\_mc} (\href{https://github.com/aragagnin/hydro_mc}{github.com/aragagnin/hydro\_mc}).
 This tool is able to perform all kinds of conversions presented in this paper and we provided a number of ready-to-use examples:  mass-concentration relation presented in Table \ref{tbl:megafit} (\linka{http://github.com/aragagnin/hydro_mc/blob/master/examples/sample_mc.py}), mass-mass conversion with fit parameters in Table \ref{tbl:megafit_mm} (\linka{http://github.com/aragagnin/hydro_mc/blob/master/examples/sample_mm.py}), and mass-mass conversion through the Mc relation in Table \ref{tbl:megafit} (\linka{http://github.com/aragagnin/hydro_mc/blob/master/examples/sample_mm_from_mc_relation.py}).
}

%
% ACKNOOOO
%

\section*{Acknowledgements}

The \textit{Magneticum Pathfinder} simulations were partially performed at the Leibniz-Rechenzentrum with CPU time assigned to the Project `pr86re'.  AR is supported by the EuroEXA project (grant no. 754337). KD acknowledges support by DAAD contract number 57396842. AR acknowledges support by  MIUR-DAAD contract number 34843  ,,The Universe in a Box''. AS and PS are supported by the ERC-StG 'ClustersXCosmo' grant agreement 716762. AS is supported by the the FARE-MIUR grant 'ClustersXEuclid' R165SBKTMA and by INFN InDark Grant. This work was supported by the Deutsche Forschungsgemeinschaft (DFG, German  Research  Foundation)  under  Germany's  Excellence Strategy - EXC-2094 - 390783311.
We are especially grateful for the support by M. Petkova through the Computational Center for Particle and Astrophysics (C$^2$PAP). Information on the \textit{Magneticum Pathfinder} project is available at \url{http://www.magneticum.org}. We acknowledge the use of \cite{Bocquet2016} python package to produce the MCMC plots. We thank the referee for the useful comments, which we believe significantly improved the clarity of the manuscript. We also thanks Stefano Andreon for the useful feedback on this paper.

\section*{Data Availability}

The data underlying this article will be shared on reasonable request to the corresponding author.

\bibliographystyle{mnras}
\bibliography{references}

\appendix

\section{Effects of baryons}
\label{ap:effects}
In this appendix section, we show the importance of correctly describing baryon physics on the estimation of halo concentration.
Since all simulations (C1-C15) share the same initial conditions, it is possible to study the evolution of the 
same halo that evolved differently in different cosmologies.

\begin{figure*}
  \caption{Evolution of virial and scale radii    and concentration of haloes in simulations C1 and C1\_norad. Upper panel shows the stacked average over $50$ haloes of ratios of $c_{\vir}$ (magenta top line), $1/r_s$ (cyan middle line) and $R_{\vir} $ (bottom green line) between  C1 and C1\_norad. Lower panel shows the evolution $R_{\vir}$ (blue top line) and $r_s$ (orange bottom line) and $c_\vir$ in blue, of  the same  halo in the simulation C1 (bottom left panels) and C1\_norad (bottom right panels).}
  \includegraphics[width=0.95\linewidth]{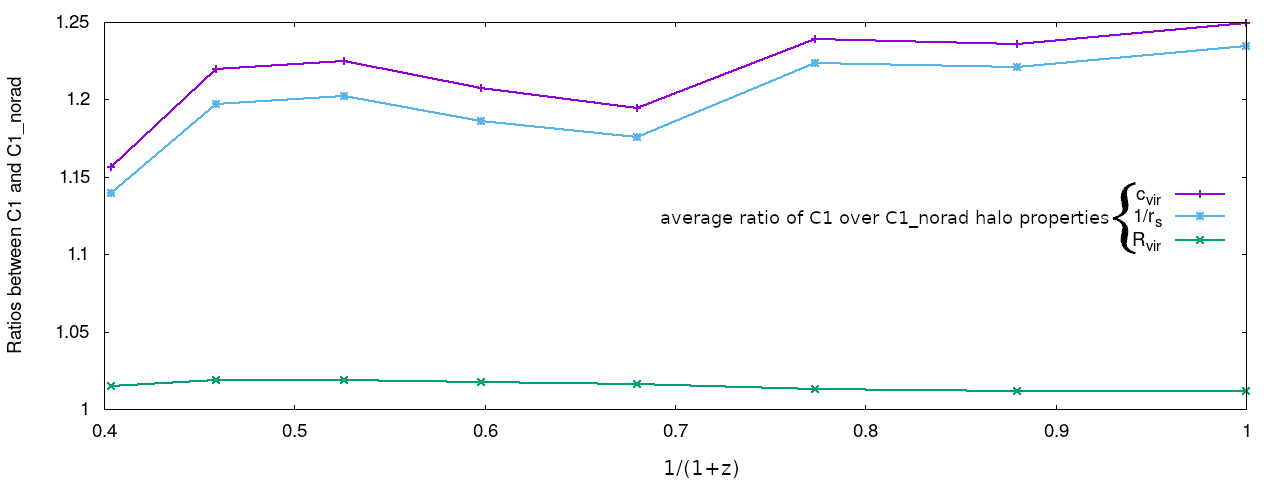}
  \includegraphics[width=0.95\linewidth]{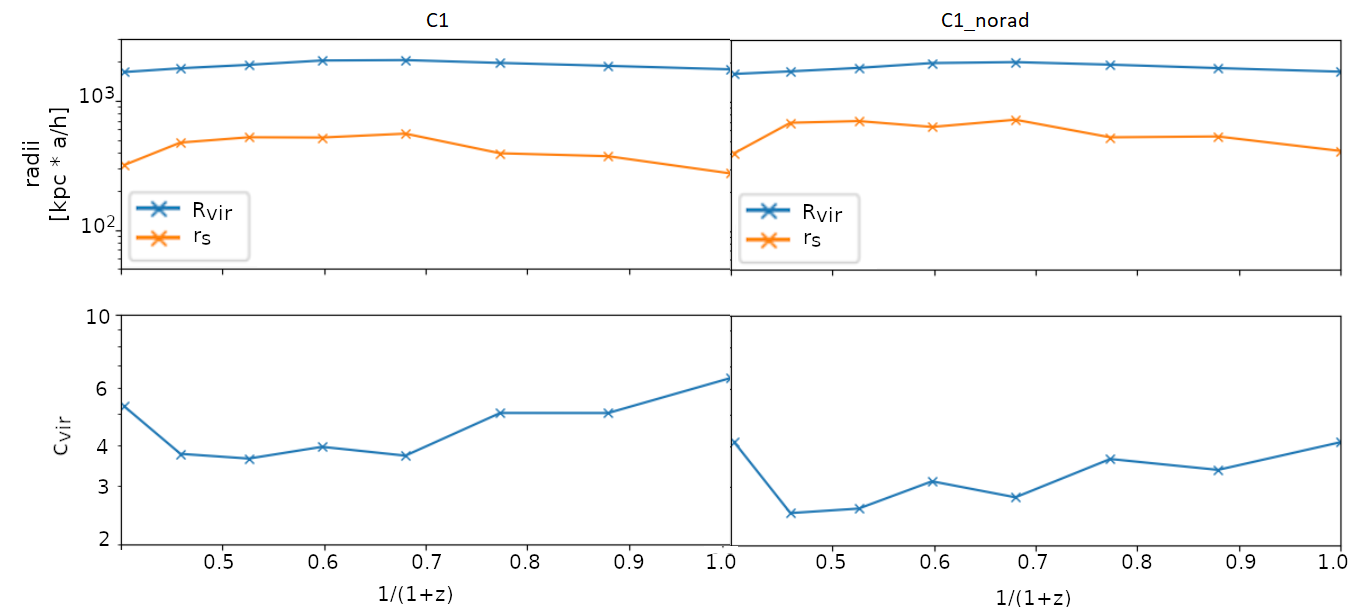}
\label{fig:c1_radnorad}
\end{figure*}

Figure \ref{fig:c1_radnorad} shows the evolution of both the virial radii and scale radii of haloes in C1 and C1\_norad.
Figure \ref{fig:c1_radnorad} (upper panel) shows the stacked ratio of concentration, virial radius and  the scale radius.
On an average, C1 haloes have higher concentration parameters ($\approx 10-15\%$ higher, up to $\approx20\%$) and this difference grows with  time.
Intuitively one may think that the difference in concentration between C1 and C1\_norad would be due to a difference in  the virial radius. However, the figure shows that it is the scale radius that produce the difference in concentration between the full physics run and the non-radiative one.

Figure \ref{fig:c1_radnorad} (bottom panel) focuses on the evolution of a single halo (bottom left panel shows the evolution of the halo in C1, whereas, the bottom right panel shows the same halo in C1\_norad).
Simulations without radiative cooling produce haloes with lower  concentration with respect to their full physics counterpart (i.e. $c_{\texttt vir}\approx 6$ lowers down to $c_{\texttt vir}\approx 5$).
This example shows that in non-radiative simulations, concentration decreases even if the full physics counterpart is characterised by the same accretion history (''jumps'' in concentration and $r_s$ values happen at the same scale factor).

\begin{figure*}
  \caption{Kinetic term $K$ $vs.$ energy from the surface pressure $E_s$ scaled by total potential energy $W$ for the same initial condition  evolved with baryon physics (left panel) and a DMO run (right panel). Black solid lines show the median $E_s/W.$}
  \includegraphics[width=\linewidth]{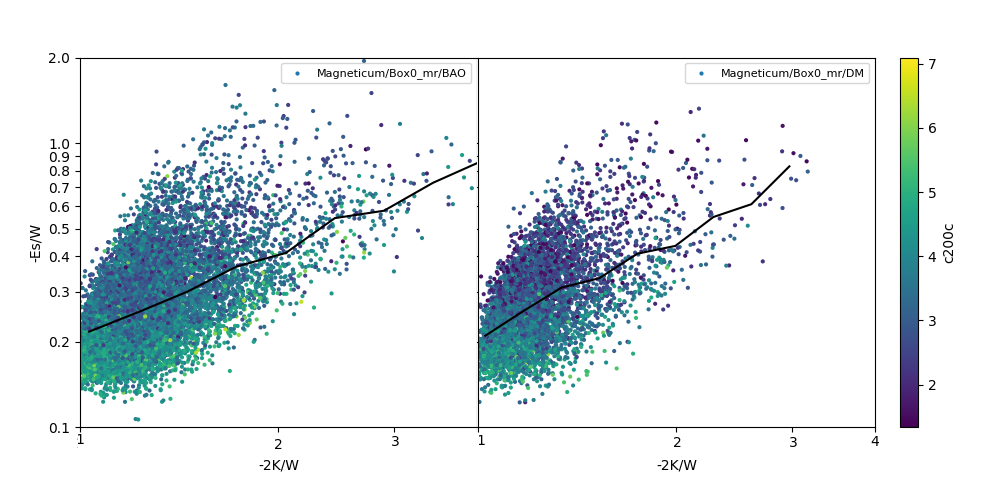}
\label{fig:esk}
\end{figure*}

\grassetto{Dynamical state is known to be related to halo concentrations~\citep{2012MNRAS.427.1322L} and  can be quantified using the virial ratio   \citep{2017MNRAS.464.2502C}, $\left(2T  - E_s\right)/W$,  where $W$ is the total potential energy, $T$  is the total kinetic energy (including gas thermal pressure) and   $E_s$ is the energy from surface pressure $P$ (from kinetic and thermal energy) at the
halo boundary. As described in \cite{1961hhs..book.....C}, $E_s$ is gievn by,}
\begin{equation}
    E_s =  \oint P(\bold r)\bold r \cdot d\bold S.
\end{equation}

\grassetto{
\cite{2017MNRAS.464.2502C} showed that baryonic physics can lower the virial ratio up to $10\%$ w.r.t. DMO runs and  \cite{2016ApJ...820...85Zhang} showed that merger timescale is shortened by a factor of up to 3 for merging clusters with gas fractions $0.15$, compared to the timescale obtained with no gas.}

\grassetto{
Figure \ref{fig:esk} shows $E_s/W$ $vs.$ $K/W$   for a DMO run (left panel) and a hydrodynamic run (right panel)   that shares the same initial conditions}\footnote{We use Magneticum/Box0\_mr simulation, with $2.7 Gpc/h$ size and gravitational softening down to $2.6\times10^9 kpc/h,$ gas and DM mass particles of $2.6 10^{9}\Msun/h$ and   $2.6 10^{10}\Msun/h$ respectively, as presented in \cite{2016MNRAS.456.2361B}.}.
\grassetto{
The two runs display a different behaviour for highly concentrated objects ($c>4$): DMO ones have low surface pressure and low total kinetic energy, while hydrodynamic ones show a much more complex and noisy relation between $Es, W$ and $c_{200c}.$}

\grassetto{It is well known that concentration does depend on dynamical state. Here we also noted how hydrodynamic simulations compared to DMO runs do show even a more noisy and complex relation between concentration and the virial ratio. 
However, given that the majority of observational studies that investigate large cluster samples lack data to accurately determine their dynamical state (see e.g. studies presented in \citealt{2015MNRAS.449..685Hoekstra}, \citealt{2016MNRAS.461.3794Okabe}, \citealt{2018A&C....24..129Melchior},  \citealt{2019MNRAS.483.2871D},  \citealt{2019MNRAS.485.4863M}, \citealt{2019ApJ...878...55B} and references therein) they will benefit from a mass-concentration relation built from hydrodynamic simulations that already averages over all possible dynamical states of a halo, as in this work.}

The average concentration of haloes shown in Figure \ref{fig:obs} are lower than the concentration computed using the dark-matter density profile presented in a previous work on Magneticum simulations~\citep[][which uses the same cosmology as C8]{2019MNRAS.486.4001R}.
The median concentration for cosmology C8 is $c_{200c}\approx3.5$ for the total matter profile, while the dark matter concentration presented in \citep{2019MNRAS.486.4001R} has $c_{200c}\approx4.3.$ 

\begin{figure}
  \caption{Density profile of both dark matter (dashed black) and total matter (dashed pink) up to the virial radius $R_{\vir}=930kpc/h$ and the corresponding NFW profile (solid lines) for a halo of C1 simulation at $z=0.$ Vertical lines correspond respectively to the dark matter profile scale radius ($139kpc/h$) and the total matter profile has a scale radius $r_s=154kpc/h.$}
  \includegraphics[width=1.1\linewidth]{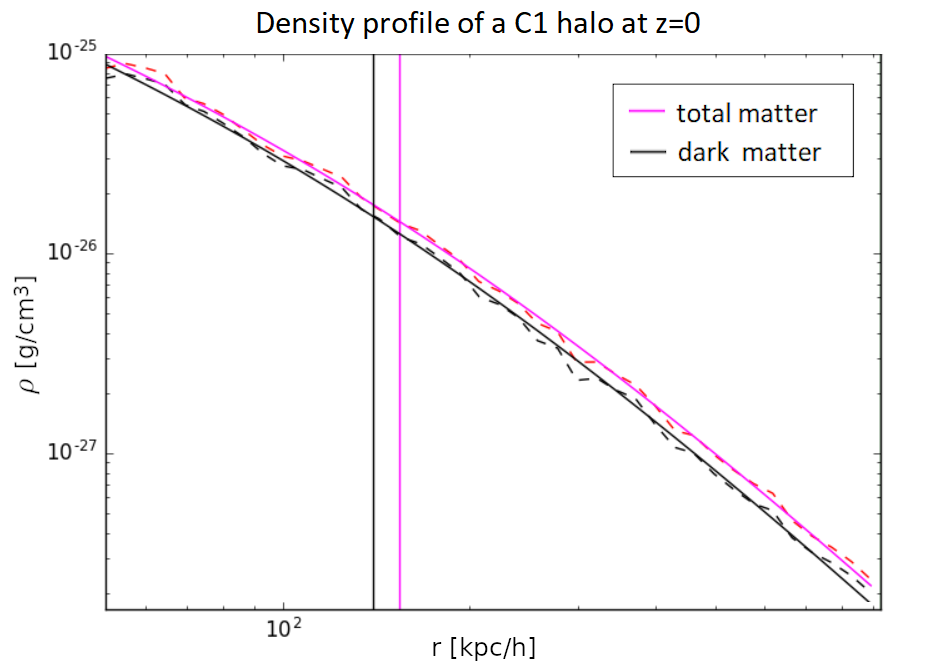}
  \label{fig:ceco}
\end{figure}

Such discrepancy is due to the fact that dark matter component is more peaked in the central region with respect to the total matter density.
Figure \ref{fig:ceco} shows an example of the matter density profiles of a Magneticum halo.
\grassetto{
Here the DM halo has a scale radius of $139$ kpc/h while the total matter scale radius is $154$ kpc/h: collisional particles and stars formed from them (and their associated heating processes as SN and AGN feedback) are capable of  lowering  a concentration parameter of $\approx20\%$.
}

\section{Cosmology-Mass-Redshift-concentration relation lite}
\label{ap:mcc}

Given the weak dependency of mass from the concentration (at least in the mass range of interests of cluster of galaxies),
we provide a cosmology-redshift-mass-concentration fit  where, in Eq. \ref{eq:mcz} we parametrise the dependency of the cosmology only in the normalisation and in the redshift dependency as the following:
\begin{equation}
\begin{split}
A &= A_0 + \alpha_m \Log \left(\frac{\Omega_m}{\Omega_{m,p}}\right) +  \alpha_b \Log \left(\frac{\Omega_b}{\Omega_{b,p}}\right)  +\\
& + \alpha_\sigma \Log \left(\frac{\sigma_8}{\sigma_{8,p}}\right)  + \alpha_h \Log \left(\frac{h_0}{h_{0,p}}\right)\\
B &= B_0\\
C &= C_0 + \gamma_m \Log \left(\frac{\Omega_m}{\Omega_{m,p}}\right) +  \gamma_b \Log \left(\frac{\Omega_b}{\Omega_{b,p}}\right)  +\\
& + \gamma_\sigma \Log \left(\frac{\sigma_8}{\sigma_{8,p}}\right)  + \gamma_h \Log \left(\frac{h_0}{h_{0,p}}\right)
\end{split}
\label{eq:abclite}
\end{equation}

%
% tbl:megafit
%
\begin{table*}
    \caption{Pivots and fit parameters for the cosmology dependent redshift-mass-concentration plane as Table \ref{tbl:megafit}, here the logarithmic slope of mass is not dependent on cosmology, thus we fit Eq. \ref{eq:mcz} and Eq. \ref{eq:abclite}, for concentration overdensities of $\Delta=\Delta_{\vir},\Delta_{200c},\Delta_{500c},\Delta_{2500c}$ and $\Delta_{200m}.$ The pivots  $\Omega_{m,p},\Omega_{b,p},\sigma _8$ and $h_0$ in Eq. \ref{eq:abc} are the cosmological parameters of C8 as in Table \ref{tbl:sims} ($\Omega_m=0.272, \Omega_b=0.0456, \sigma_8=0.809, h_0=0.704$).  Errors on $A_0,B_0,C_0$ and $\sigma$ are omitted as they are all $<0.001\%.$  The package \texttt{hydro\_mc} contains a script that utilises this relation (\linka{http://github.com/aragagnin/hydro_mc/blob/master/examples/sample_mc_lite.py}).}
    \begin{tabular}{llllll}
\hline
  \multirow{2}{*}{Parameter}&\multicolumn{5}{c}{Overdensity}\\
                 & vir                     & 200c                    & 500c                    & 2500c                   & 200m                    \\
\hline
$M_p [M_{\odot}]$   & $1.99e+14$              & $1.74e+14$              & $1.37e+14$              & $6.87e+13$              & $2.24e+14$              \\
 $a_p$           & $0.877$                 & $0.877$                 & $0.877$                 & $0.877$                 & $0.877$                 \\
 $A_0$           & $1.499 $       & $1.238 $       & $0.859 $       & $0.122 $       & $1.688 $       \\
 $B_0$           & $-0.048 $      & $-0.053 $      & $-0.060 $      & $-0.037 $      & $-0.044 $      \\
 $C_0$           & $0.520 $       & $0.201 $       & $0.187 $       & $0.110 $       & $0.910 $       \\
 $\alpha_m$      & $0.423 \pm 0.006$       & $0.60 \pm 0.01$         & $0.63 \pm 0.01$         & $0.7273 \pm 0.0006$     & $0.201 \pm 0.003$       \\
 $\alpha_b$      & $-0.141 \pm 0.006$      & $-0.152 \pm 0.006$      & $-0.131 \pm 0.005$      & $-0.179 \pm 0.004$      & $-0.186 \pm 0.006$      \\
 $\alpha_\sigma$ & $0.65 \pm 0.02$         & $0.65 \pm 0.02$         & $0.61 \pm 0.03$         & $0.516 \pm 0.003$       & $0.60 \pm 0.02$         \\
 $\alpha_h$      & $-0.28 \pm 0.01$        & $-0.25 \pm 0.02$        & $-0.27 \pm 0.02$        & $-0.23 \pm 0.01$        & $-0.17 \pm 0.02$        \\
 $\gamma_m$      & $0.19 \pm 0.01$         & $0.360 \pm 0.010$       & $0.336 \pm 0.009$       & $0.36 \pm 0.01$         & $-0.10 \pm 0.01$        \\
 $\gamma_b$      & $0.02 \pm 0.06$         & $-0.15 \pm 0.06$        & $-0.04 \pm 0.05$        & $0.00 \pm 0.07$         & $0.00 \pm 0.06$         \\
 $\gamma_\sigma$ & $0.76 \pm 0.05$         & $0.72 \pm 0.04$         & $0.89 \pm 0.04$         & $0.94 \pm 0.06$         & $0.61 \pm 0.05$         \\
 $\gamma_h$      & $-0.4 \pm 0.2$          & $-0.1 \pm 0.2$          & $-0.4 \pm 0.2$          & $-0.5 \pm 0.2$          & $-0.4 \pm 0.2$          \\
 $\sigma$        & $0.388031 $ & $0.384516 $ & $0.376690 $ & $0.382868 $ & $0.388477 $ \\
\hline
\end{tabular}
    \label{tbl:megafit_lite}
\end{table*}

Table \ref{tbl:megafit_lite} show the results of this fit, with the same procedure as in Section \ref{sec:mcac}, where pivot values are the ones for the reference cosmology C8 and errors are assigned by performing the same fit as in \cite{2019Singh}.

%
% tbl:megafit_dm
%
\begin{table*}[ht]
    \caption{Fit parameters for the cosmology dependent redshift-mass-concentration plane as Table \ref{tbl:megafit}, here we computed the concentration using the scale radius of the dark matter density profile, plus the logarithmic slope of mass is not dependent on cosmology. We fit Eq. \ref{eq:mcz} and Eq. \ref{eq:abclite}, for concentration overdensities of $\Delta=\Delta_{\vir},\Delta_{200c},\Delta_{500c},\Delta_{2500c}$ and $\Delta_{200m}.$  The pivots  $\Omega_{m,p},\Omega_{b,p},\sigma_8$ and $h_0$ in Eq. \ref{eq:abc} are the cosmological parameters of C8 as in Table \ref{tbl:sims} ($\Omega_m=0.272, \Omega_b=0.0456, \sigma_8=0.809, h_0=0.704$).  Errors on $A_0,B_0,C_0$ and $\sigma$ are omitted as they are all $<0.001\%.$  The package \texttt{hydro\_mc} contains a script that utilises this relation (\linka{http://github.com/aragagnin/hydro_mc/blob/master/examples/sample_mc_dm_lite.py}).}
    \begin{tabular}{rlllll}
\hline
  \multirow{2}{*}{Parameter}&\multicolumn{5}{c}{Overdensity}\\

                & vir                & 200c                    & 500c               & 2500c                   & 200m                    \\
\hline
 $A_0$           & $1.499 $  & $1.238 $       & $0.979 $  & $0.213 $       & $1.798 $       \\
 $B_0$           & $-0.048 $ & $-0.053 $      & $-0.039 $ & $-0.015 $      & $-0.034 $      \\
 $C_0$           & $0.520 $  & $0.201 $       & $0.178 $  & $0.055 $       & $0.918 $       \\
$\alpha_m$      & $0.42 \pm 0.05$    & $0.60 \pm 0.01$         & $0.46 \pm 0.07$    & $0.588 \pm 0.001$       & $0.008 \pm 0.007$       \\
 $\alpha_b$      & $-0.14 \pm 0.03$   & $-0.152 \pm 0.006$      & $-0.08 \pm 0.03$   & $-0.204 \pm 0.010$      & $-0.072 \pm 0.006$      \\
 $\alpha_\sigma$ & $0.65 \pm 0.03$    & $0.65 \pm 0.02$         & $0.47 \pm 0.05$    & $0.363 \pm 0.006$       & $0.53 \pm 0.01$         \\
 $\alpha_h$      & $-0.28 \pm 0.05$   & $-0.25 \pm 0.02$        & $-0.33 \pm 0.05$   & $-0.47 \pm 0.03$        & $0.03 \pm 0.01$         \\
 $\gamma_m$      & $0.19 \pm 0.04$    & $0.360 \pm 0.010$       & $0.34 \pm 0.01$    & $0.51 \pm 0.03$         & $-0.23 \pm 0.01$        \\
 $\gamma_b$      & $0.02 \pm 0.06$    & $-0.15 \pm 0.06$        & $-0.4 \pm 0.1$     & $-0.7 \pm 0.1$          & $-0.09 \pm 0.06$        \\
 $\gamma_\sigma$ & $0.76 \pm 0.06$    & $0.72 \pm 0.04$         & $0.5 \pm 0.1$      & $0.3 \pm 0.1$           & $0.45 \pm 0.02$         \\
 $\gamma_h$      & $-0.4 \pm 0.2$     & $-0.1 \pm 0.2$          & $-1.1 \pm 0.4$     & $-1.9 \pm 0.5$          & $0.02 \pm 0.06$         \\
 $\sigma$        & $0.39 $    & $0.384516 $ & $0.51 $    & $0.484290 $ & $0.498887 $ \\
\hline
\end{tabular}
    \label{tbl:megafit_dm_lite}
\end{table*}

Table \ref{tbl:megafit_dm_lite} show the results of the mass-concentration plane where we  fit the NFW profile of the dark matter density profile only.
The functional form is as in Eq. \ref{eq:abclite}, with the same procedure as the previous one (thus, as  in Section \ref{sec:mcac}).

\section{Theoretical scatter of Mass conversion using an Mc relation}
\label{ap:scat}

Equation system \ref{eq:mm}   shows how the    concentration in an overdensity $\Delta_2$ is uniquely identified by the concentration in $\Delta_1$ by solving bottom equation in Eq. \ref{eq:mm}.
Although there are four variables in Eq.  \ref{eq:mm} (namely $M_\du$, $M_\dd$, $c_\du$ and $c_\dd$), since there are two equations the system  depends on two of them.

 \cite{2003ApJ...584..702H}  provides a fitting formula for $c_{\Delta2}$ as a function of  $c_{\Delta1}.$
 On the other hand since $c_{\Delta2}$ depends   monotonically  from right side of Eq. \ref{eq:mm}, in this work we convert the values from $c_{\Delta1}$ to $c_{\Delta2}$ using the fixed-point technique derived by solving equation \ref{eq:mm} the Banach-Cacioppoli theorem \citep[see e.g.][for a review]{ciesielski2007}.

To evaluate $c_{\Delta2}$ we start with a guess value of  $c_{\Delta1}$ and iteratively apply it to Eq.  \ref{eq:mm} in order to get the new value of value of $c_{\Delta2}$, until it converges, practically we fix $\frac{\Delta_1}{\Delta_2}$ and $c_{\Delta,1}$  rewrite Eq. \ref{eq:mm} as

\begin{equation}
    \begin{split}
    \tilde c\left(x\right) &\equiv& c_{\Delta1} \cdot\left(\frac{\Delta_1}{\Delta_2} \frac{f(x)}{f(c_{\Delta1})}\right)^{\frac{1}{3}}\\
    c_{\Delta2}& =&  \tilde c\left(c_{\Delta2}\right).
    \end{split}
    \label{eq:bcs}
\end{equation}

We found that the relative error after $9$ iterations is, at the worst, comparable with \cite{2003ApJ...584..702H} and can go down to $10^{-9}$ for concentration values higher than $20.$
As a first value we choose $c_{\Delta 1},$ so

\begin{equation}
    c_{\Delta2} \approx     \tilde c\left( \tilde c\left( \tilde c\left( \tilde c\left( \tilde c\left( \tilde c\left( \tilde c\left( \tilde c\left( \tilde c\left( c_{\Delta1}\right)\right)\right)\right)\right)\right)\right)\right)\right)
    \label{eq:c9}
\end{equation}

 \begin{figure}
  \caption{Relative error when converting the concentration using Eq. \ref{eq:bcs} (i.e. Banach-Cacioppoli theorem) or using the method proposed in \protect\cite{2003ApJ...584..702H} }
  \includegraphics[width=\linewidth]{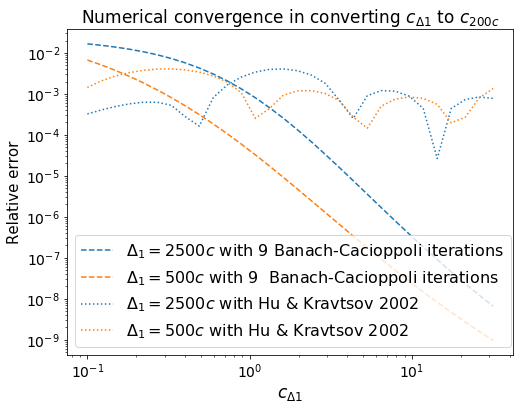}
  \label{fig:c_converg}
\end{figure}

Figure \ref{fig:c_converg} shows the relative error when converting $M_{500c}$ and $M_{2500c}$ to $M_{200c}.$ Both approach have an error smaller than $\approx 0.1\%,$ while the iteration proposed here can reach much more precise value and it is easier to implement.  Only $9$ iterations  produce a relative error that in the worst case is comparable with technique in  \cite{2003ApJ...584..702H} and it is capable of going down to $10^{-8}.$

 \begin{figure}
  \caption{Analytical uncertainty on the concentration obtained by the theoretical propagation of error}
  \includegraphics[width=\linewidth]{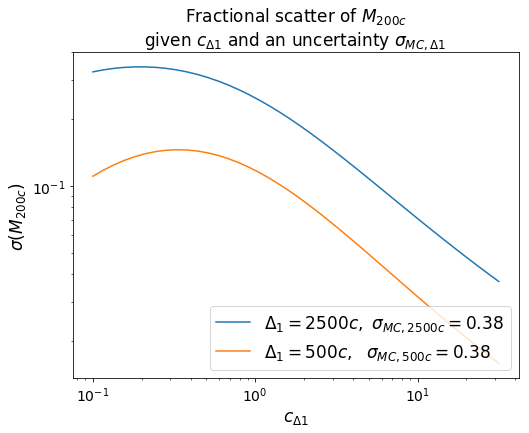}
  \label{fig:c_theo}
\end{figure}

Figure \ref{fig:c_theo} show the convsersion from overdensities $\Delta_2 = 2500$ and $\Delta_2 = 500$ to $\Delta_1 = 200.$ These relations are nearly linear with a deviation for lower concentrations.

Another interesting property of Eq. \ref{eq:mm} is the possibility of knowing $M_{\Delta2}/M_{\Delta1}$ only by knowing $c_{\Delta1}.$

 \begin{figure}
  \caption{Analytical value of $c_{200c}$   for a given concentration $c_{\Delta1}.$ We used $\Delta1=500c$ and  $\Delta1=2500c.$ }
  \includegraphics[width=\linewidth]{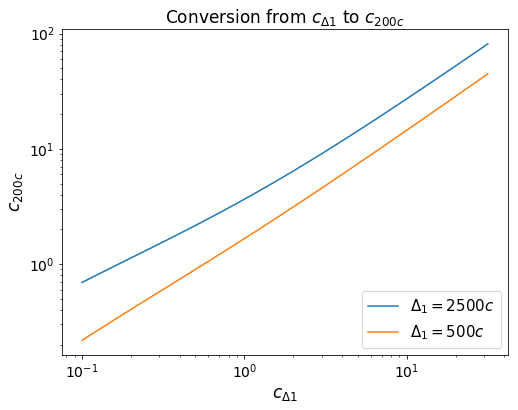}
 \label{fig:c_frac}
\end{figure}

Figure \ref{fig:c_frac} shows such conversions for overdensitites $\Delta_{2500c}$ and $\Delta_{500c}$ to $\Delta_{200c}1.$
This conversion gets flatter and flatter as the concentration increases, implying that the higher the concentration the lower the error one makes in this conversion.

It is possible to estimate this uncertainty analytically. Given the fact that Mc relations are nown with uncertainties, it is interesting to see how to propagate the error analitically when converting from $c_\du$ to $c_\dd,$ which is proportional to the derivarive caming from Eq. \ref{eq:m_delta}:

\begin{equation}\begin{split}
    \frac{dc_\dd}{dc_\du} &=& \frac{c_\dd}{c_\du} + \frac{1}{3}
    \frac{c_\dd}{f\left(c_\du\right)}\frac{df(c)}{dc}\bigg\rvert_{c=c_\du}\frac{dc_\dd}{dc_\du} -\\
    &&-\frac{1}{3}
    \frac{c_\dd}{f\left(c_\du\right)}\frac{df(c)}{dc}\bigg\rvert_{c=c_\dd},
    \end{split}
\label{eq:dcdc_1}
\end{equation}
where $f(c)$ is, in case of imposing a NFW profile, given in Eq. \ref{eq:fc}.
One can rearrange Eq. \ref{eq:dcdc_1}  to isolate the derivative:
\begin{equation}
    \cfrac{dc_\dd}{dc_\du} = \cfrac{ \cfrac{c_\dd}{c_\du} - \cfrac{1}{3}
    \cfrac{c_\dd}{f\left(c_\du\right)}\cfrac{df(c)}{dc}\bigg\rvert_{c=c\dd}   }{
   1-\cfrac{1}{3}
    \cfrac{c_\dd}{f\left(c_\du\right)}\cfrac{df(c)}{dc}\bigg\rvert_{c=c\du}
    }
\label{eq:dcdc}
\end{equation}

One can understand how a uncertainty propagates analytically from $M_\dd\left(M_\du,c_\du\right)$ in Eq. \ref{eq:mm}, by computing the derivative
$$\frac{dM_\dd}{dc_\du} = \frac{\partial M_\dd}{\partial c_\du} + \frac{\partial M_\dd}{\partial M_\du} \frac{dM_\du}{dc_\du},  $$
given the very weak dependency of mass from concentration, we can approximate
$$ \frac{dM_\du}{dc_\du} \approx 0,$$
one gets
$$ \frac{dM_\dd}{dc_\du} = 3M_\dd\left(\frac{1}{c_\dd}\frac{dc_\dd}{dc_\du}-\frac{1}{c_\du}\right),   $$
where $dc_\dd / dc_\du$ is evaluated as in Eq. \ref{eq:dcdc}.

 \begin{figure}
  \caption{Analytical value of $M_{200c}$ with respect to $M_{\Delta1}$ for a given concentration $c_{\Delta1}.$ We used $\Delta1=500c$ and  $\Delta1=2500c.$ }
  \includegraphics[width=\linewidth]{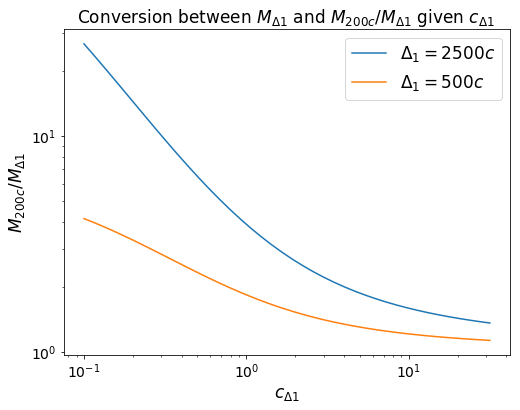}
  \label{fig:c_prop}
\end{figure}

Figure \ref{fig:c_prop} show the uncertainty variation when converting to $M_{200}$ for a scatter in the concentration compatible with the scatter we found in our Mc relation (see Table \ref{tbl:megafit}). This is helpful in understanding the actual scatter one find in real case scenarios as Sections \ref{seq:mmconv} and \ref{seq:mmconvmc}.

% Don't change these lines
\bsp	% typesetting comment
\label{lastpage}
\end{document}